\documentclass[%
 aip,pof,
 amsmath,amssymb,
 reprint,%
author-year,%
]{revtex4-1}
\usepackage{graphicx}
\usepackage{dcolumn}
\usepackage{bm}

\usepackage[utf8]{inputenc}
\usepackage[T1]{fontenc}
\usepackage{mathptmx}
\usepackage{etoolbox}

\usepackage{amsthm}
\usepackage{mathtools}
\usepackage{cancel}
\usepackage{subcaption}
\usepackage{todonotes}

\makeatletter
\def\@email#1#2{%
 \endgroup
 \patchcmd{\titleblock@produce}
  {\frontmatter@RRAPformat}
  {\frontmatter@RRAPformat{\produce@RRAP{*#1\href{mailto:#2}{#2}}}\frontmatter@RRAPformat}
  {}{}
}%
\makeatother
\begin{document}
\preprint{0}

\title{Shape evolution of fluid deformable surfaces under active geometric forces}
\author{Maik Porrmann}
\affiliation{Institute of Scientific Computing, TU Dresden, 01062 Dresden, Germany}

\author{Axel Voigt}%
\affiliation{Institute of Scientific Computing, TU Dresden, 01062 Dresden, Germany}
\affiliation{Center for Systems Biology Dresden, Pfotenhauerstr. 108, 01307 Dresden, Germany}
\affiliation{Cluster of Excellence, Physics of Life, TU Dresden, 01062 Dresden, Germany}
\email{axel.voigt@tu-dresden.de, corresponding author: Axel Voigt}

\date{\today}

\begin{abstract}
Models for fluid deformable surfaces provide valid theories to describe the dynamics of thin fluidic sheets of soft materials. To use such models in morphogenesis and development requires to incorporate active forces. We consider active geometric forces which respond to mean curvature gradients. Due to these forces perturbations in shape can induce tangential flows which can enhance the perturbation leading to shape instabilities. We numerically explore these shape instabilities and analyse the resulting dynamics of closed surfaces with constant enclosed volume. The numerical approach considers surface finite elements and a semi-implicit time stepping scheme and shows convergence properties, similar to those proven to be optimal for Stokes flow on stationary surfaces.
\end{abstract}

\maketitle

\newcommand{\R}{\mathbb{R}}
\newcommand{\C}{\mathbb{C}}
\newcommand{\N}{\mathbb{N}}

\renewcommand{\vec}[1]{\bm{#1}}
\newcommand{\surf}{\mathbb{S}}
\newcommand{\shape}{\bm{B}}
\newcommand{\gaussCurv}{\mathbb{K}}
\newcommand{\proj}{\bm{P}}

\newcommand{\covGrad}{\nabla_{\surf}}
\newcommand{\covDiv}{\operatorname{div}_\surf}
\newcommand{\gradS}{\covGrad}
\newcommand{\gradC}{\nabla_{C}}
\newcommand{\divC}{\operatorname{div}_C}

\newcommand{\gradP}{\nabla_{\proj}}

\newcommand{\restrict}[2]{ #1|_{ #2 }}
\newcommand{\cof}{\bm{Cof}}
\newcommand{\sym}{\operatorname{sym}}
\newcommand{\norm}[1]{\lvert \! \lvert #1 \rvert \! \rvert}

\newcommand{\DiscreteSpace}[2]{\mathcal{V}^{#1,#2}_{h}}

\newcommand{\eg}{e.\,g.}

\newcommand{\Ac}{\mathrm{Ac}}
\newcommand{\Be}{\mathrm{Be}}
\newcommand{\Rey}{\Reynolds}
\newcommand{\LR}{\mathrm{LR}}
\newcommand{\Vred}{v_{red}}
\newcommand{\InnerProd}[2]{\left( #1,#2 \right) _h}

\theoremstyle{remark}
\newtheorem{thm}{Theorem}
\newtheorem{prob}[thm]{Problem}
\newtheorem{lem}[thm]{Lemma}
\newtheorem{prop}[thm]{Proposition}
\newtheorem{remark}[thm]{Remark}
\renewcommand{\S}{\mathcal{S}}
\renewcommand{\H}{\mathcal{H}}
\newcommand{\K}{\mathcal{K}}
\newcommand{\B}{\mathcal{B}}
\newcommand{\red}[1]{{\color{red}#1}}
\newcommand{\Id}{\textbf{Id}}
\renewcommand{\div}{\text{div}}
\newcommand{\Reyn}{\text{Re}}
\newcommand{\surfupd}{\partial_{\bm{X}} }
\newcommand{\paramUpdate}{W}
\newcommand{\LScalar}[2]{\left\langle #1,#2 \right\rangle_h}
\newcommand{\Tr}{\text{Tr}}


\newcommand{\Reylightian}{\mathcal{R}}

\newcommand{\ra}[1]{\renewcommand{\arraystretch}{#1}}

\newtheorem{problem}{Problem}
\newtheorem{example}{Example}

\newcommand{\T}{\mbox{\tiny T}}
\newcommand{\disp}{\mbox{\tiny disp}}
\newcommand{\Forall}{\forall \,}

\newcommand{\timeSymbol}{t}
\newcommand{\Time}{\timeSymbol} 
\newcommand{\TimeF}{T} 
\newcommand{\tempo}{\Time} 
\newcommand{\Tempo}{\TimeF} 

\newcommand{\VolForm}{\Diff V_{\First}}

\newcommand{\FirstSymbol}{\mathcal{G}}
\newcommand{\First}[1][]{
  \ifthenelse{\equal{#1}{}}
  {\FirstSymbol}
  {\FirstSymbol_{\scriptscriptstyle{#1}}}
}
\newcommand{\EigFirst}{{\mu_{\scriptscriptstyle{\First}}}}
\newcommand{\EigFirstInvmin}{g_{*}}
\newcommand{\EigFirstInvmax}{g^*}
\newcommand{\CminDetG}{c_{\scriptscriptstyle{\First}}}
\newcommand{\CmaxDetG}{C_{\scriptscriptstyle{\First}}}
\newcommand{\ConstGeom}[1][]{
   \ifthenelse{\equal{#1}{}}
   {K_{\First}\,}
   {K_{\First,#1}\,}
}

\newcommand{\FirstForm}[1][]{
  \ifthenelse{\equal{#1}{}}
  {\operatorname{I_{\point}}}
  {\operatorname{I_{#1}}}
}
\newcommand{\GradSymbol}{\operatorname{\mathbf{\nabla}}}
\newcommand{\Grad}{\GradSymbol}
\newcommand{\Div}{\Grad\cdot}
\newcommand{\Lap}{\Delta}

\newcommand{\GradSurf}{\GradSymbol_{\!\SurfDomain}\!}
\newcommand{\divS}{\operatorname{div}_{\! \SurfDomain}\!}

\newcommand{\GradP}{\GradSymbol_{\!\ProjMat}\!}
\newcommand{\DivP}{\operatorname{div}_{\!\ProjMat}\!}

\newcommand{\GradC}{\GradSymbol_{\!C}\!}
\newcommand{\DivC}{\operatorname{div}_{\!C}\!}

\newcommand{\LapSurf}{\Lap_{\SurfDomain}}
\newcommand{\GradSurfH}{\GradSymbol_{\SurfDomain[\meshparam]}}
\newcommand{\DivSurfH}{\GradSurfH\cdot}
\newcommand{\LapSurfH}{\Lap_{\SurfDomain[\meshparam]}}

\newcommand{\GradSurfConv}[1]{\GradSymbol_{#1}}

\newcommand{\GradSurfcoord}[1]{\underline{D}_{#1}}

\newcommand{\gradDef}[1]{\eth_{#1}}
\newcommand{\gradDefMat}[1]{\gradDef{#1}^{\SurfDomain}}
\newcommand{\gradDefLower}[1]{\gradDef{#1}^{b}}

\newcommand{\gradFunc}[1]{\mathbf{D}_{#1}}

\newcommand{\Der}[1][]
{
  \ifthenelse{\equal{#1}{}}
  {\partial}
  {\partial_{\scriptscriptstyle{#1}}}
}
\newcommand{\DerT}{\Der_{\Time}\,}
\newcommand{\DerPar}[2]{\frac{\Der #1}{\Der #2}}
\newcommand{\tDerPar}[2]{\tfrac{\Der #1}{\Der #2}}
\newcommand{\DerParT}[1]{\DerPar{#1}{\Time}}
\newcommand{\DerParIL}[2]{{\Der #1}/{\Der #2}}
\newcommand{\DerTot}[2][t]
{
  \ifthenelse{\equal{#2}{}}
  {\frac{d #2}{dt}}
  {\frac{d #2}{d #1}}
}
\newcommand{\Derd}[2][]{
  \ifthenelse{\equal{#1}{}}
  {\Diffsymbol{#2}}
  {\Diffsymbol_{{#1}}{#2}}
}
\newcommand{\Diffsymbol}{\operatorname{d}\!}
\newcommand{\Diff}[2][]{
  \ifthenelse{\equal{#1}{}}
  {\Diffsymbol{#2}}
  {\Diffsymbol{#2}_{{#1}}}
}
\newcommand{\DirDerSymb}{D}
\newcommand{\DirDer}[2][]
{
  \ifthenelse{\equal{#1}{}}
  {\DirDerSymb^{#2}}
  {\DirDerSymb^{#2}_{#1}}
}

\newcommand{\tr}{\operatorname{tr}}

\newcommand{\Graph}[1]{\operatorname{Graph}(#1)}

\newcommand{\BigO}[1]{\ensuremath{\operatorname{\mathcal{O}}\!\left(#1\right)}}
\newcommand{\LittleO}[1]{\ensuremath{\operatorname{\mathpzc{o}}\!\left(#1\right)}}
\newcommand{\Rank}[1]{\ensuremath{\operatorname{rank}\left(#1\right)}}
\newcommand{\Exp}[1]{\ensuremath{\operatorname{e}^{#1}}}
\newcommand{\ExpIL}[1]{\ensuremath{\operatorname{exp}({#1})}}
\newcommand{\REALsymbol}{\mathbb{R}}
\newcommand{\REAL}[1][]{
  \ifthenelse{\equal{#1}{}}
  {\REALsymbol}
  {{\REALsymbol}^{#1}}
}
\newcommand{\EUCLsymbol}{\mathbb E}
\newcommand{\EUCL}[1][]{
  \ifthenelse{\equal{#1}{}}
  {\EUCLsymbol}
  {{\EUCLsymbol}^{#1}}
}
\newcommand{\NATURALsymbol}{\mathbb N}
\newcommand{\NATURAL}[1][]{
  \ifthenelse{\equal{#1}{}}
  {\NATURALsymbol}
  {{\NATURALsymbol}^{#1}}
}

\newcommand{\dist}[2]{\operatorname{dist}(#1,#2)}

\newcommand{\ABS}[2][]
{
  \ifthenelse{\equal{#1}{}}
  {\left| #2 \right|}
  {\left| #2 \right|_{#1}}
}

\newcommand{\ABSSurf}[1]{\ensuremath{\left|#1\right|_{\SurfDomain}}}

\newcommand{\NORM}[2][]
{
  \ifthenelse{\equal{#1}{}}
  {\left\| #2 \right\|}
  {\left\| #2 \right\|_{#1}}
}
\newcommand{\vertiii}[1]{{\left\vert\kern-0.25ex\left\vert\kern-0.25ex\left\vert #1 \right\vert\kern-0.25ex\right\vert\kern-0.25ex\right\vert}}
\newcommand{\BrNORM}[2][]
{
  \ifthenelse{\equal{#1}{}}
  {\vertiii{#2}}
  {\vertiii{#2}_{#1}}
}

\newcommand{\SCAL}[3][]
{
  \ifthenelse{\equal{#1}{}}
  {\left\langle{#2},{#3}\right\rangle}
  {\left\langle{#2},{#3}\right\rangle_{#1}}
}
\newcommand{\SCALF}[3][]
{
  \ifthenelse{\equal{#1}{}}
  {\left({#2},{#3}\right)}
  {\left({#2},{#3}\right)_{#1}}
}
\newcommand{\CSCALF}[2]{{C}\left(#1,#2\right)}
\newcommand{\BSCALF}[2]{b\left(#1,#2\right)}
\newcommand{\BSCALFH}[2]{b_h\left(#1,#2\right)}
\newcommand{\DSCALF}[2]{d\left(#1,#2\right)}
\newcommand{\scalar}[2]{\prodscal{#1}{#2}}
\newcommand{\scalprod}[2]{\left\langle #1,#2 \right\rangle}
\newcommand{\scalprodSurf}[3][]
{
  \ifthenelse{\equal{#1}{}}
  {\left\langle {#2},{#3} \right\rangle_{\scriptscriptstyle\SurfDomain}}
  {\left\langle {#2},{#3} \right\rangle_{{#1}}}
}
\newcommand{\Lie}[2]{\left[#1,#2\right]}
\newcommand{\DET}[1]{\ensuremath{\operatorname{det}(#1)}}
\newcommand{\Trace}[1]{\ensuremath{\operatorname{Tr}(#1)}}
\newcommand{\Inner}[2]{\left({#1}\,,\,{#2}\right)_{\SurfDomain}}
\newcommand{\InnerApprox}[2]{\left({#1}\,,\,{#2}\right)_{\meshparam}}

\newcommand{\Origin}{\mathbf{o}} 
\newcommand{\Point}{\mathbf{P}}
\newcommand{\point}[1][]
{
  \ifthenelse{\equal{#1}{}}
  {\mathbf{p}}
  {\mathbf{p}_{#1}}
}
\newcommand{\Qpoint}{\mathbf{Q}}
\newcommand{\qpoint}{\mathbf{q}}
\newcommand{\qvalue}{q}
\newcommand{\rpoint}{\mathbf{r}}
\newcommand{\mpoint}{\mathbf{m}}
\newcommand{\midPoint}{\mathbf{m}}
\newcommand{\lform}[1]{{f}\left(#1\right)}
\newcommand{\Lform}[1]{{L}\left(#1\right)}
\newcommand{\Poly}{\mathcal{P}}
\newcommand{\RT}{\mathcal{RT}}
\newcommand{\Const}{\mathcal{C}}

\newcommand{\Dt}{\Delta\tempo}
\newcommand{\Dx}{\Delta x}

\newcommand{\From}{:}
\newcommand{\To}{\longrightarrow}
\newcommand{\Closure}[1]{\operatorname{cl}({#1})}

\newcommand{\Domain}{\Omega}
\newcommand{\Bnd}{\partial \Omega}
\newcommand{\RegionSymb}{R}
\newcommand{\Region}[1][]
{
  \ifthenelse{\equal{#1}{}}
  {\RegionSymb}
  {\RegionSymb_{#1}}
}
\newcommand{\PartitionRegion}{\mathcal{R}}
\newcommand{\diam}{\operatorname{diam}}
\newcommand{\diamReg}{\diam\Region}
\newcommand{\IT}{[0,\TimeF]} 
\newcommand{\Interval}[1][]
{
  \ifthenelse{\equal{#1}{}}
  {I}
  {I_{#1}}
}
\newcommand{\tInterval}{\tilde{I}}
\newcommand{\SurfDomainsymb}{\mathcal{S}}
\newcommand{\SurfDomain}[1][]{
  \ifthenelse{\equal{#1}{}}
  {\SurfDomainsymb}
  {{\SurfDomainsymb_{#1}}}
}
\newcommand{\tSurfDomain}[1][]{
  \ifthenelse{\equal{#1}{}}
  {\tilde{\SurfDomainsymb}}
  {\tilde{\SurfDomainsymb}_{#1}}
}
\newcommand{\SurfDomainBndsymb}{\partial\Gamma}
\newcommand{\SurfDomainBnd}[1][]{
  \ifthenelse{\equal{#1}{}}
  {\SurfDomainBndsymb}
  {\SurfDomainBndsymb_{#1}}
}
\newcommand{\ClosedSurfDomain}[1][]{
  \ifthenelse{\equal{#1}{}}
  {\Closedsymb{\SurfDomain}}
  {\Closedsymb{\SurfDomain[#1]}}
}

\newcommand{\curve}{\sigma}
\newcommand{\tcurve}{\tilde{\curve}}
\newcommand{\DerCurve}{{\curve}'}
\newcommand{\tDerCurve}{{\tcurve}'}
\newcommand{\curveparam}{\lambda}
\newcommand{\bcurveparam}{\tau}

\newcommand{\heightsymb}{\mathcal{H}}
\newcommand{\height}[1][]{
  \ifthenelse{\equal{#1}{}}
  {\heightsymb}
  {\heightsymb_{#1}}
}
\newcommand{\FSMsymbol}{\mathcal{F}}
\newcommand{\BSMsymbol}{\mathcal{B}}
\newcommand{\LSMsymbol}{\mathcal{L}}
\newcommand{\SSMsymbol}{\mathcal{M}}
\newcommand{\FSM}
  {\FSMsymbol}
\newcommand{\BSM}[1][]
{
  \ifthenelse{\equal{#1}{}}
  {\BSMsymbol}
  {\BSMsymbol_{#1}}
}
\newcommand{\LSM}
  {\LSMsymbol}
\newcommand{\SSM}
  {\SSMsymbol}
\newcommand{\FreeWaterSurfHat}{\mathcal{S}_{\hat{\FSMsymbol}}}
\newcommand{\FreeWaterSurf}{\mathcal{S}_\FSMsymbol}
\newcommand{\Bottom}{\mathcal{S}_\BSMsymbol}
\newcommand{\Lateral}{\mathcal{S}_\LSMsymbol}
\newcommand{\Surf}{\mathcal{S}}
\newcommand{\BoundSurf}{\SSMsymbol}
\newcommand{\SurfBndsymb}{\partial\Surf}
\newcommand{\SurfBnd}[1][]{
  \ifthenelse{\equal{#1}{}}
  {\SurfBndsymb}
  {\SurfBndsymb_{#1}}
}
\newcommand{\Closedsymb}[1]{\bar{#1}}
\newcommand{\ClosedSurf}[1][]{
  \ifthenelse{\equal{#1}{}}
  {\Closedsymb{\Surf}}
  {\Closedsymb{\Surf[#1]}}
}

\newcommand{\Vector}[1]{\mathbf{#1}}

\newcommand{\press}{p}
\newcommand{\tpress}{\tilde{\press}}
\newcommand{\grav}{g}
\newcommand{\gravpot}{\zcg}
\newcommand{\vectgrav}{\mathbf{g}}
\newcommand{\Depth}{\eta}
\newcommand{\tension}{\sigma}
\newcommand{\ttension}{\tilde{\tension}}
\newcommand{\densitySymb}{\rho}
\newcommand{\density}[1][]{
  \ifthenelse{\equal{#1}{}}
  {\densitySymb}
  {\densitySymb_{\scriptscriptstyle{#1}}}
}

\newcommand{\eb}{\boldsymbol{e}} 

\newcommand{\concSymbol}{\phi}
\newcommand{\CHsol}{\concSymbol}
\newcommand{\CHsolApprox}{\CHsol_{\meshparam}}
\newcommand{\pressApprox}{\press_{\meshparam}}

\newcommand{\dWellSymbol}{W}
\newcommand{\dWell}{\dWellSymbol}
\newcommand{\mobilitySymbol}{m}
\newcommand{\mobility}{\mobilitySymbol}
\newcommand{\Galpha}{{G_{\alpha}}}
\newcommand{\chempotSub}{\mu_{\scriptscriptstyle{CH}}}
\newcommand{\chempot}{\mu}
\newcommand{\chempotApprox}{\mu_{\meshparam}}

\newcommand{\bendingSymb}{\kappa}
\newcommand{\bendStiff}[1][]{
  \ifthenelse{\equal{#1}{}}
  {\bendingSymb}
  {\bendingSymb_{#1}}
}

\newcommand{\bendStiffGauss}[1][]{
  \ifthenelse{\equal{#1}{}}
  {\overline{\bendingSymb}}
  {\overline{\bendingSymb}_{#1}}
}

\newcommand{\Reynolds}{\operatorname{Re}}

\newcommand{\idvec}[1]{\boldsymbol{e}_{#1}} 
\newcommand{\idvecT}[1]{\boldsymbol{e}_{#1}^T} 

\newcommand{\AspRatio}{\epsilon}
\newcommand{\epsCurv}{\AspRatio_{\scriptscriptstyle \First}}
\newcommand{\epsApprox}{\epsCurv}
\newcommand{\interfaceparam}{\AspRatio}

\newcommand{\param}{\boldsymbol{X}}

\newcommand{\update}{\boldsymbol{Y}}
\newcommand{\updateApprox}{\update_{\meshparam}}

\newcommand{\MapUsymb}{\param}
\newcommand{\MapU}[1][]
{
  \ifthenelse{\equal{#1}{}}
    {\MapUsymb}
    {\MapUsymb_{#1}}
}
\newcommand{\MapLinsymb}{F}
\newcommand{\MapLin}[1][]
{
  \ifthenelse{\equal{#1}{}}
  {\ensuremath{\MapLinsymb}}
  {\ensuremath{\MapLinsymb_{#1}}}
}

\newcommand{\MapVsymb}{\psi}
\newcommand{\MapV}[1][]
{
  \ifthenelse{\equal{#1}{}}
    {\MapVsymb}
    {\MapVsymb_{#1}}
}
\newcommand{\Transsymb}{\Phi}
\newcommand{\Trans}[1][]
{
  \ifthenelse{\equal{#1}{}}
    {\Transsymb}
    {\Transsymb_{\scriptscriptstyle{#1}}}
}
\newcommand{\InvMapsymb}{\Psi}
\newcommand{\InvMap}[1][]
{
  \ifthenelse{\equal{#1}{}}
    {\InvMapsymb}
    {\InvMapsymb_{\scriptscriptstyle{#1}}}
}

\newcommand{\fsymb}{f}
\newcommand{\scalFun}[1][]
{
  \ifthenelse{\equal{#1}{}}
  {\fsymb}
  {\fsymb_{#1}}
}
\newcommand{\tscalFun}[1][]
{
  \ifthenelse{\equal{#1}{}}
  {\tilde{\fsymb}}
  {\tilde{\fsymb}_{#1}}
}
\newcommand{\bscalFun}[1][]
{
  \ifthenelse{\equal{#1}{}}
  {\bar{\fsymb}}
  {\bar{\fsymb}_{#1}}
}
\newcommand{\gsymb}{g}
\newcommand{\scalFung}[1][]
{
  \ifthenelse{\equal{#1}{}}
  {\gsymb}
  {\gsymb_{#1}}
}
\newcommand{\Fsymb}{F}
\newcommand{\FvecFun}[1][]
{
  \ifthenelse{\equal{#1}{}}
  {\Fsymb}
  {\Fsymb_{#1}}
}
\newcommand{\tFvecFun}[1][]
{
  \ifthenelse{\equal{#1}{}}
  {\tilde{\Fsymb}}
  {\tilde{\Fsymb}_{#1}}
}
\newcommand{\Ffunc}[2][]
{
  \ifthenelse{\equal{#1}{}}
  {\ensuremath{\Fsymb_{#2}}}
  {\ensuremath{\Fsymb^{#1}_{#2}}}
}
\newcommand{\Gsymb}{g}
\newcommand{\Gfun}[1][]
{
  \ifthenelse{\equal{#1}{}}
  {\ensuremath{\Gsymb}}
  {\ensuremath{\Gsymb_{{#1}}}}
}
\newcommand{\bGfun}[1][]
{
  \ifthenelse{\equal{#1}{}}
  {\ensuremath{\bar{\Gsymb}}}
  {\ensuremath{\bar{\Gsymb}_{{#1}}}}
}
\newcommand{\bGfunApprox}{\bGfun[\meshparam]}
\newcommand{\preimagesymb}{{-1}} 
\newcommand{\preimage}[2]{#1^\preimagesymb(#2)}
\newcommand{\inverse}[1]{#1^{-1}}

\newcommand{\Jac}{\mathbf{J}}
\newcommand{\Hess}{\mathbf{H}}
\newcommand{\PrincipalK}[1][]
{
  \ifthenelse{\equal{#1}{}}
  {k}
  {k_{#1}}
}

\newcommand{\vecsymb}{u}
\newcommand{\vecFun}[1][]{
  \ifthenelse{\equal{#1}{}}
  {\Vector{\vecsymb}}
  {\vecsymb^{#1}}
}
\newcommand{\tvecFun}[1][]{
  \ifthenelse{\equal{#1}{}}
  {\tilde{\vecsymb}}
  {\tilde{\vecsymb}_{#1}}
}

\newcommand{\wwsymb}{w}
\newcommand{\ww}[1][]
{
  \ifthenelse{\equal{#1}{}}
  {\mathbf{\wwsymb}}
  {\wwsymb^{#1}}
}
\newcommand{\uusymb}{u}
\newcommand{\uu}[1][]
{
  \ifthenelse{\equal{#1}{}}
  {\mathbf{\uusymb}}
  {\uusymb^{#1}}
}
\newcommand{\positVec}{\mathbf{r}}

\newcommand{\VecFieldSymbol}{X}
\newcommand{\VecField}[1][]
{
  \ifthenelse{\equal{#1}{}}
  {\VecFieldSymbol}
  {\VecFieldSymbol^{#1}}
}
\newcommand{\VecFieldSymbolC}{Y}
\newcommand{\VecFieldYSymbol}{\boldsymbol{\VecFieldSymbolC}}
\newcommand{\VecFieldY}[1][]
{
  \ifthenelse{\equal{#1}{}}
  {\VecFieldYSymbol}
  {\VecFieldYSymbol^{#1}}
}
\newcommand{\VecFieldYC}{\VecFieldSymbolC}

\newcommand{\xvsymb}{x}
\newcommand{\xv}[1][]
{
  \ifthenelse{\equal{#1}{}}
  {\mathbf{\xvsymb}}
  {\mathbf{\xvsymb}_{\scriptscriptstyle{#1}}}
}
\newcommand{\xvcomp}[1][]{
  \ifthenelse{\equal{#1}{}}
  {\xvsymb}
  {\xvsymb^{\scriptscriptstyle{#1}}}
}
\newcommand{\xcg}[1][]{
  \ifthenelse{\equal{#1}{}}
  {\xvcomp[1]}
  {\xvcomp[1]_{\scriptscriptstyle{#1}}}
}
\newcommand{\ycg}[1][]{
  \ifthenelse{\equal{#1}{}}
  {\xvcomp[2]}
  {\xvcomp[2]_{\scriptscriptstyle{#1}}}
}
\newcommand{\zcg}[1][]{
  \ifthenelse{\equal{#1}{}}
  {\xvcomp[3]}
  {\xvcomp[3]_{\scriptscriptstyle{#1}}}
}
\newcommand{\gcscomp}{\xcg,\ycg,\zcg}

\newcommand{\Reftrianglesymb}{\zeta}
\newcommand{\Refcomp}[1]{\Reftrianglesymb^{#1}}
\newcommand{\Refx}{\Reftrianglesymb^1}
\newcommand{\Refy}{\Reftrianglesymb^2}
\newcommand{\zv}{\mathbf{\Reftrianglesymb}}

\newcommand{\svsymb}{s}
\newcommand{\sv}[1][]{
  \ifthenelse{\equal{#1}{}}
  {\mathbf{\svsymb}}
  {\mathbf{\svsymb}_{\scriptscriptstyle{#1}}}
}
\newcommand{\svcomp}[1][]
{
  \ifthenelse{\equal{#1}{}}
  {\svsymb}
  {\svsymb^{\scriptscriptstyle{#1}}}
}
\newcommand{\xcl}[1][]{
  \ifthenelse{\equal{#1}{}}
   {\svcomp[1]}
   {\svcomp[1]_{\scriptscriptstyle{#1}}}
}
\newcommand{\ycl}[1][]{
  \ifthenelse{\equal{#1}{}}
   {\svcomp[2]}
   {\svcomp[2]_{\scriptscriptstyle{#1}}}
}
\newcommand{\zcl}[1][]{
  \ifthenelse{\equal{#1}{}}
   {\svcomp[3]}
   {\svcomp[3]_{\scriptscriptstyle{#1}}}
}
\newcommand{\lcscomp}{\xcl,\ycl,\zcl}

\newcommand{\ProjSymb}{\operatorname{\pi}}
\newcommand{\ProjFun}[2][]{
  \ifthenelse{\equal{#1}{}}
  {\ProjSymb\left(#2\right)}
  {\ProjSymb_{\scriptscriptstyle{#1}}\left(#2\right)}
}
\newcommand{\Prm}[1][]
{
  \ifthenelse{\equal{#1}{}}
  {\operatorname{pr}}
  {\operatorname{pr}_{\scriptscriptstyle{#1}}}
}
\newcommand{\TanPlane}[2][]
{
  \ifthenelse{\equal{#1}{}}
  {T_{\scriptscriptstyle{\point}}#2}
  {T_{\scriptscriptstyle{#1}}#2}
}
\newcommand{\Tbundle}[1]{\operatorname{T}{#1}}
\newcommand{\Vbundle}{\operatorname{E}}

\newcommand{\SubsetSymbol}{\mathcal{U}}
\renewcommand{\Subset}{\SubsetSymbol}
\newcommand{\SubsetU}[1][]
{
  \ifthenelse{\equal{#1}{}}
  {{U}}
  {{U}_{#1}}
}
\newcommand{\SubsetV}[1][]
{
  \ifthenelse{\equal{#1}{}}
  {{V}}
  {{V}_{#1}}
}
\newcommand{\SubsetW}[1][]
{
  \ifthenelse{\equal{#1}{}}
  {{W}}
  {{W}_{#1}}
}
\newcommand{\NeighSymbol}{\mathcal{N}}
\newcommand{\Neigh}[1][]
{
  \ifthenelse{\equal{#1}{}}
  {\NeighSymbol_{\point}}
  {\NeighSymbol_{#1}}
}
\newcommand{\NeighSurf}[1][]
{
  \ifthenelse{\equal{#1}{}}
  {\SubsetSymbol_{\point}}
  {\SubsetSymbol_{#1}}
}
\newcommand{\NormSymb}{\nu}
\newcommand{\normalvec}[1][]
{
  \ifthenelse{\equal{#1}{}}
  {\boldsymbol{\NormSymb}}
  {\NormSymb_{#1}}
}\newcommand{\normalvecApprox}{\NormSymb_{\meshparam}}
\newcommand{\normalSurf}[1][]
{
  \ifthenelse{\equal{#1}{}}
  {\NormSymb}
  {\NormSymb(#1)}
}
\newcommand{\normalInterp}[1][]
{
  \ifthenelse{\equal{#1}{}}
   {\tilde{\NormSymb}}
   {\tilde{\NormSymb}_{\scriptscriptstyle{#1}}}
}

\newcommand{\normalEdge}{\mathbf{\nu}}
\newcommand{\basisCC}{t}
\newcommand{\basisGC}{e}
\newcommand{\vecBaseGC}[1][]
{
  \ifthenelse{\equal{#1}{}}
  {\mathbf{\basisGC}}
  {\mathbf{\basisGC}_{#1}}
}
\newcommand{\vecBasePhys}[1][]
{
  \ifthenelse{\equal{#1}{}}
  {\mathbf{\basisGC}}
  {\mathbf{\basisGC}_{#1}}
}

\newcommand{\vecBaseCCcv}[1][]
{
  \ifthenelse{\equal{#1}{}}
  {\mathbf{\basisCC}}
  {\mathbf{\basisCC}_{#1}}
}
\newcommand{\tangential}{\vecBaseCCcv[1],\vecBaseCCcv[2],\vecBaseCCcv[3]}

\newcommand{\tvecBaseCCcv}[1][]
{
  \ifthenelse{\equal{#1}{}}
  {\tilde{\mathbf{\basisCC}}}
  {\tilde{\mathbf{\basisCC}}_{#1}}
}
\newcommand{\hvecBaseCCcv}[1][]
{
  \ifthenelse{\equal{#1}{}}
  {\hat{\mathbf{\basisCC}}}
  {\hat{\mathbf{\basisCC}}_{#1}}
}
\newcommand{\vecBaseCCctrv}[1][]
{
  \ifthenelse{\equal{#1}{}}
  {\mathbf{\basisCC}}
  {\mathbf{\basisCC}^{#1}}
}

\newcommand{\metricsymbol}{g}
\newcommand{\metrTensCv}[1]{\metricsymbol_{\mbox{\tiny{#1}}}}
\newcommand{\hatmetrTensCv}[1]{\hat{\metricsymbol}_{\mbox{\tiny{#1}}}}
\newcommand{\metrTensCtrv}[1]{\metricsymbol^{#1}}
\newcommand{\metrcoefsymbol}{h}
\newcommand{\metrcoef}[1]{\metrcoefsymbol_{\mbox{\tiny{(#1)}}}}
\newcommand{\metrcoefDef}[1]{
  \sqrt{
    \left(\frac{\Der\xcg}{\Der\sv_{#1}}\right)^2
   +\left(\frac{\Der\ycg}{\Der\sv_{#1}}\right)^2
   +\left(\frac{\Der\zcg}{\Der\sv_{#1}}\right)^2}
 }
\newcommand{\defVecBaseCCcv}[1]
{
  \left(
    \dfrac{\Der\xcg[]}{\Der\svcomp[#1]},
    \dfrac{\Der\ycg[]}{\Der\svcomp[#1]},
    \dfrac{\Der\zcg[]}{\Der\svcomp[#1]}
  \right)
}

\newcommand{\first}[1]{
  \IfEqCase{#1}{
    {1}{\operatorname{E}}
    {2}{\operatorname{F}}
    {3}{\operatorname{G}}
  }
  [\PackageError{first}{Undefined option to first: #1}{}]%
}
\newcommand{\SecondFormSymbol}{\ensuremath{\operatorname{II}}}
\newcommand{\SecondForm}[1][]
{
  \ifthenelse{\equal{#1}{}}
  {\SecondFormSymbol_{\point}}
  {\SecondFormSymbol_{#1}}
}
\newcommand{\second}[1]{
  \IfEqCase{#1}{
    {1}{\operatorname{e}}
    {2}{\operatorname{f}}
    {3}{\operatorname{g}}
  }
  [\PackageError{first}{Undefined option to first: #1}{}]
}
\newcommand{\WeigSymbol}{\mathcal{W}}
\newcommand{\Weig}[1][]
{
  \ifthenelse{\equal{#1}{}}
  {\WeigSymbol}
  {\WeigSymbol_{#1}}
}
\newcommand{\ChristSymb}[2]{\Gamma_{#1}^{#2}}

\newcommand{\face}{{\sigma}}
\newcommand{\parvar}{\sigma}

\newcommand{\DivEins}[2][]{
  {\dfrac{\Der {#2}}{\Der #1}}
}

\newcommand{\velSymbol}{\boldsymbol{u}}
\newcommand{\vectvel}[1][]
{
   \ifthenelse{\equal{#1}{}}
   {\mathbf{\velSymbol}}
   {\mathbf{\velSymbol}(#1)}
}
\newcommand{\velVSymbol}{\boldsymbol{w}}
\newcommand{\vectvelV}[1][]
{
   \ifthenelse{\equal{#1}{}}
   {\mathbf{\velVSymbol}}
   {\mathbf{\velVSymbol}(#1)}
}
\newcommand{\vectvelApprox}[1][]
{
   \ifthenelse{\equal{#1}{}}
   {\mathbf{\velSymbol}_{\meshparam}}
   {\mathbf{\velSymbol}_{\meshparam}(#1)}
}
\newcommand{\redvectvel}{
  \underline{
     \vec{\velSymbol}
     }
}
\newcommand{\velcompContr}[2][i]
{
   \ifthenelse{\equal{#2}{}}
   {\velSymbol^{#1}}
   {\velSymbol^{#1}(#2)}}
\newcommand{\velcompPhys}[2][i]
{
   \ifthenelse{\equal{#2}{}}
   {\velSymbol_{(#1)}}
   {\velSymbol_{(#1)}(#2)}}
\newcommand{\velcomp}{\velcompContr}

\newcommand{\velSymbolRP}{v}
\newcommand{\velcompContrRP}[2][i]
{
   \ifthenelse{\equal{#2}{}}
   {\velSymbolRP^{#1}}
   {\velSymbolRP^{#1}(#2)}}
\newcommand{\velRP}[1][]
{
  \ifthenelse{\equal{#1}{}}
  {\velSymbolRP}
  {\velSymbolRP_{#1}}
}
\newcommand{\velcompRP}{\velRP}

\newcommand{\velcompApprox}[2][i]
{
   \ifthenelse{\equal{#2}{}}
   {\velSymbol^{#1)}}
   {\velSymbol^{#1}_{(#2)}}
}

\newcommand{\coord}{\mathbf{x}}
\newcommand{\coordC}{x}

\newcommand{\VelSymbol}{U}
\newcommand{\vectVel}[1][]
{
   \ifthenelse{\equal{#1}{}}
   {\vec{\VelSymbol}}
   {\vec{\VelSymbol}(#1)}
}
\newcommand{\Velcomp}[2][i]
{
   \ifthenelse{\equal{#2}{}}
   {\VelSymbol^{#1}}
   {\VelSymbol^{#1}(#2)}
}
\newcommand{\VprimoSymbol}{\tilde{u}}
\newcommand{\Vprimo}[1][]
{
   \ifthenelse{\equal{#1}{}}
   {\VprimoSymbol}
   {\VprimoSymbol(#1)}
}
\newcommand{\VprimoComp}[2][i]
{
   \ifthenelse{\equal{#2}{}}
   {\VprimoSymbol^{#1}}
   {\VprimoSymbol^{#1}(#2)}
}
\newcommand{\ttvelSymbol}{\tilde{\Mymathbb{u}}}
\newcommand{\ttvel}[1][]
{
   \ifthenelse{\equal{#1}{}}
   {\mathbf{\ttvelSymbol}}
   {\mathbf{\ttvelSymbol}(#1)}
}
\newcommand{\ttvelComp}[2][i]
{
   \ifthenelse{\equal{#2}{}}
   {\ttvelSymbol^{#1}}
   {\ttvelSymbol^{#1}(#2)}
}

\newcommand{\alphaSymbol}{\alpha}
\newcommand{\alphacomp}[1]{\alphaSymbol_{#1}}
\newcommand{\MatAlphaSymbol}{\mathbb{A}}
\newcommand{\MatAlpha}[1][]{%
  \ifthenelse{\equal{#1}{}}
  {\MatAlphaSymbol}
  {\MatAlphaSymbol_{#1}}
}

\newcommand{\QSymbol}{q}
\newcommand{\Qdisch}[1][]
{
   \ifthenelse{\equal{#1}{}}
   {\mathbf{\QSymbol}}
   {\mathbf{\QSymbol}(#1)}
}
\newcommand{\Qcomp}[2][i]
{
   \ifthenelse{\equal{#2}{}}
   {\QSymbol^{#1}}
   {\QSymbol^{#1}(#2)}
}
\newcommand{\Qvect}[1][]
{
   \ifthenelse{\equal{#1}{}}
   {\mathbf{\QSymbol}}
   {\mathbf{\QSymbol}(#1)}
}
\newcommand{\FricSymbol}{f}
\newcommand{\vectFric}[1][]
{
   \ifthenelse{\equal{#1}{}}
   {\mathbf{\FricSymbol}}
   {\mathbf{\FricSymbol}_{\scriptscriptstyle{#1}}}
}
\newcommand{\Friccomp}[2][i]
{
   \ifthenelse{\equal{#2}{}}
   {\FricSymbol_{#1}}
   {\FricSymbol_{#1}(#2)}}
\newcommand{\BFsymbol}{\tau}
\newcommand{\BottomFriction}[1][]{
  \ifthenelse{\equal{#1}{}}
  {\BFsymbol_{b}}
  {\BFsymbol_{b}^{#1}}
}

\newcommand{\ProjMatSymb}{\boldsymbol{P}}
\newcommand{\ProjMat}[1][]{
  \ifthenelse{\equal{#1}{}}
  {\ProjMatSymb}
  {\ProjMatSymb_{#1}}
}
\newcommand{\IDSymbol}{\boldsymbol{I}}
\newcommand{\IDtens}[1][]{
  \ifthenelse{\equal{#1}{}}
  {\IDSymbol}
  {\IDSymbol(#1)}
}
\newcommand{\IDMat}{\IDSymbol}
\newcommand{\Onevect}{\Mymathbb{1}}

\newcommand{\tensSymbol}{\boldsymbol{T}}
\newcommand{\tenscompSymbol}{\tau}
\newcommand{\tens}[1][]{
  \ifthenelse{\equal{#1}{}}
  {\tensSymbol}
  {\tensSymbol(#1)}
}
\newcommand{\tenscomp}[2][ij]
{
  \ifthenelse{\equal{#2}{}}
  {\tenscompSymbol^{#1}}
  {\tenscompSymbol^{#1}(#2)}
}
\newcommand{\tensrow}[2][i]
{
  \ifthenelse{\equal{#2}{}}
  {\tensSymbol^{(#1)}}
  {\tensSymbol^{(#1)}(#2)}
}
\newcommand{\TensSymbol}{\mathbf{T}}
\newcommand{\Tens}[1][]{
  \ifthenelse{\equal{#1}{}}
  {\TensSymbol}
  {\TensSymbol_{#1}}
}

\newcommand{\TensCompSymbol}{\TensSymbol}
\newcommand{\TensComp}[2][ij]
{
  \ifthenelse{\equal{#2}{}}
  {\TensCompSymbol^{#1}}
  {\TensCompSymbol^{#1}(#2)}
}

\newcommand{\tensPrimoSymbol}{\tilde{\mathbf{\tau}}}
\newcommand{\tensPrimo}[1][]{
  \ifthenelse{\equal{#1}{}}
  {\tensPrimoSymbol}
  {\tensPrimoSymbol(#1)}
}
\newcommand{\tensPrimoCompSymbol}{\tensPrimoSymbol}
\newcommand{\tensPrimoComp}[2][ij]
{
  \ifthenelse{\equal{#2}{}}
  {\tensPrimoCompSymbol^{#1}}
  {\tensPrimoCompSymbol^{#1}(#2)}
}
\newcommand{\MCxl}[1][]{\ifthenelse{\equal{#1}{}}{h_{(1)}}{h_{(1),#1}}}
\newcommand{\MCyl}[1][]{\ifthenelse{\equal{#1}{}}{h_{(2)}}{h_{(2),#1}}}
\newcommand{\MCzl}[1][]{\ifthenelse{\equal{#1}{}}{h_{(3)}}{h_{(3),#1}}}
\newcommand{\DzgDxl}{\dfrac{\Der\zcg}{\Der\xcl}}
\newcommand{\DzgDyl}{\dfrac{\Der\zcg}{\Der\ycl}}
\newcommand{\DzgDzl}{\dfrac{\Der\zcg}{\Der\zcl}}
\newcommand{\DbottDxl}{\dfrac{\Der \BSM}{\Der \xcl}}
\newcommand{\DbottDyl}{\dfrac{\Der \BSM}{\Der \ycl}}
\newcommand{\DbottDzl}{\dfrac{\Der \BSM}{\Der \zcl}}

\newcommand{\MPsymb}{h}
\newcommand{\Lev}{\ell}
\newcommand{\smallestH}[1][]
{
  \ifthenelse{\equal{#1}{}}
    {{l}}
    {{l}_{#1}}
}
\newcommand{\meshparam}[1][]
{
  \ifthenelse{\equal{#1}{}}
    {\MPsymb}
    {\MPsymb_{\scriptscriptstyle{#1}}}
}
\newcommand{\InradiusSymbol}{r}
\newcommand{\Inradius}[1][]
{
  \ifthenelse{\equal{#1}{}}
  {\InradiusSymbol}
  {\InradiusSymbol_{\scriptscriptstyle{#1}}}
}
\newcommand{\Tsymb}{\mathcal{T}}
\newcommand{\Triang}[1][]
{
  \ifthenelse{\equal{#1}{}}
    {\Tsymb}
    {\Tsymb_{#1}}
}
\newcommand{\TriangH}[1][]
{
  \ifthenelse{\equal{#1}{}}
    {\Tsymb_{\meshparam}}
    {\Tsymb_{#1}}
}
\renewcommand{\Triang}{\TriangH}
\newcommand{\Edgesymb}{\sigma}
\newcommand{\Edge}[1][]{
  \ifthenelse{\equal{#1}{}}
    {\Edgesymb}
    {\Edgesymb_{#1}}
}
\newcommand{\EdgeH}[1][]{
  \ifthenelse{\equal{#1}{}}
    {\Edgesymb_{\meshparam}}
    {\Edgesymb_{\meshparam,#1}}
}
\newcommand{\NEdge}[1][]{
  \ifthenelse{\equal{#1}{}}
    {N_{\Edgesymb}}
    {N_{\Edgesymb({#1})}}
}
\newcommand{\Cellsymb}{T}
\newcommand{\Cell}[1][]{
  \ifthenelse{\equal{#1}{}}
    {\Cellsymb}
    {\Cellsymb_{#1}}
}
\newcommand{\tCell}[1][]{
  \ifthenelse{\equal{#1}{}}
    {\tilde{\Cellsymb}}
    {\tilde{\Cellsymb}_{#1}}
}
\newcommand{\CellH}[1][]{
  \ifthenelse{\equal{#1}{}}
    {\Cellsymb_{\meshparam}}
    {\Cellsymb_{\meshparam,#1}}
}
\newcommand{\areaSymb}{\mathcal{A}}
\newcommand{\CellArea}[1][]
{
  \ifthenelse{\equal{#1}{}}
    {\areaSymb_{\Cell}}
    {\areaSymb_{#1}}
}
\newcommand{\CellHArea}[1][]
{
  \ifthenelse{\equal{#1}{}}
    {\areaSymb_{\CellH}}
    {\areaSymb_{\meshparam,#1}}
}

\newcommand{\NCell}[1][]{
  \ifthenelse{\equal{#1}{}}
    {N_{\Cellsymb}}
    {N_{\Cellsymb({#1})}}
}
\newcommand{\GeodCurve}{\mathbf{c}}
\newcommand{\lengthSymb}{l}
\newcommand{\interfaceLength}{\lengthSymb}
\newcommand{\edgeLength}[1][]
{
  \ifthenelse{\equal{#1}{}}
  {\lengthSymb_{\Edge}}
  {\lengthSymb_{#1}}
}
\newcommand{\arcLength}{\edgeLength}
\newcommand{\edgeHLength}[1][]
{
  \ifthenelse{\equal{#1}{}}
  {\lengthSymb_{\EdgeH}}
  {\lengthSymb_{\meshparam,#1}}
}
\newcommand{\subVolume}[2]{V_{#1}^{#2}}

\newcommand{\Sourcesymb}{\mathbf{S}}
\newcommand{\Source}[1][]
{
  \ifthenelse{\equal{#1}{}}
    {\Sourcesymb}
    {\Sourcesymb_{#1}}
}
\newcommand{\SourceDepth}{S^{\Depth}}
\newcommand{\SourceQ}{\Sourcesymb^{\Qdisch}}
\newcommand{\SourceTermsMC}[1]{S^{\mbox{\tiny{M}}}_{#1}}
\newcommand{\SourceTermsForces}[1]{S^{\mbox{\tiny{F}}}_{#1}}
\newcommand{\SourceEdge}[1][]{
  \ifthenelse{\equal{#1}{}}
    {\Sourcesymb_{ij}}
    {\Sourcesymb_{#1}}
}

\newcommand{\ConservVar}[2]{\mathbf{U}_{#1}^{#2}}
\newcommand{\ConservVarRP}[2]{\mathbf{V}_{#1}^{#2}}
\newcommand{\tConservVar}[2]{\tilde{\mathbf{U}}_{#1}^{#2}}
\newcommand{\Flux}{\underline{\underline{F}}}
\newcommand{\FluxDepth}{\underline{F}^{\Depth}}
\newcommand{\FluxQ}{\underline{\underline{F}}^{\Qdisch}}
\newcommand{\FluxEdgesymb}{\mathbf{F}}
\newcommand{\FluxEdge}[1][]{
  \ifthenelse{\equal{#1}{}}
    {\FluxEdgesymb_{ij}}
    {\FluxEdgesymb_{#1}}
}
\newcommand{\FluxVec}{\mathbf{F}}
\newcommand{\numFlux}[1][]
{
  \ifthenelse{\equal{#1}{}}
  {\tilde{\fsymb}}
  {\tilde{\fsymb}_{#1}}
}
\newcommand{\WBflux}{\mathcal{F}}
\newcommand{\tFluxEdge}[1]{\tilde{\mathbf{F}}^{#1}_{ij}}
\newcommand{\FluxFuncX}{\mathbf{F_1}}
\newcommand{\FluxFuncY}{\mathbf{F_2}}
\newcommand{\FluxFuncNormSymbol}{\mathbf{F}}
\newcommand{\FluxFuncNorm}[1][]{
  \ifthenelse{\equal{#1}{}}
    {\FluxFuncNormSymbol^{\normalEdge}}
    {\FluxFuncNormSymbol^{\normalEdge}_{#1}}
}
\newcommand{\FluxFuncRP}[1]{\mathbf{F}_{#1}}
\newcommand{\JacobianSymbol}{\mathbf{A}}
\newcommand{\Jacobian}[1][]{
  \ifthenelse{\equal{#1}{}}
    {\JacobianSymbol}
    {\JacobianSymbol_{#1}}
}
\newcommand{\EValSymbol}{\lambda}
\newcommand{\EVal}[1][]{
  \ifthenelse{\equal{#1}{}}
  {\EValSymbol}
  {\EValSymbol_{#1}}
}
\newcommand{\EVecSymbol}{\mathbf{r}}
\newcommand{\EVec}[2][]{
  \ifthenelse{\equal{#1}{}}
  {\EVecSymbol^{(#2)}}
  {\EVecSymbol^{(#2)}_{#1}}
}

\newcommand{\Sect}[1]{S_{#1}}

\newcommand{\nodeA}{A}
\newcommand{\nodeB}{B}
\newcommand{\nodeC}{C}
\newcommand{\nodeAfem}{\point[1]}
\newcommand{\nodeBfem}{\point[2]}
\newcommand{\nodeCfem}{\point[3]}
\newcommand{\nodeDfem}{\point[4]}
\newcommand{\midPointEdge}[1][]
{
  \ifthenelse{\equal{#1}{}}
  {\midPoint_{\scriptscriptstyle\Edge}}
  {\midPoint_{\scriptscriptstyle\Edge[#1]}}
}
\newcommand{\gpPointEdgeDG}[1][]
{
  \ifthenelse{\equal{#1}{}}
  {\point_{\scriptscriptstyle\Edge}}
  {\point_{\scriptscriptstyle\Edge,#1}}
}
\newcommand{\midPointCell}[1][]
{
  \ifthenelse{\equal{#1}{}}
  {\midPoint_{\scriptscriptstyle\Cell}}
  {\midPoint_{\scriptscriptstyle\Cell[#1]}}
}
\newcommand{\tangentCurve}[1]{\tau_{#1}}
\newcommand{\tangentCurveApprox}{\tilde{\tau}_{\scriptscriptstyle{\midPointEdge}}}
\newcommand{\normalSurfApprox}{\normalInterp[\midPointEdge]}
\newcommand{\normalEdgeApprox}{\tilde{\normalEdge}_{\scriptscriptstyle{\midPointEdge}}}
\newcommand{\TanPlaneApprox}{\TanPlane[\midPointEdge]{\SurfDomain}}
\newcommand{\metricApproxEdge}{\tilde{\First}_{\midPointEdge}}
\newcommand{\metricApproxCell}{\tilde{\First}_{\midPointCell}}

\newcommand{\Err}{E}
\newcommand{\Eoc}{\mbox{eoc}_{\Lev}}

\newcommand{\energySymbol}{\mathcal{F}}
\newcommand{\energy}[1][]
{
  \ifthenelse{\equal{#1}{}}
  {\energySymbol}
  {\energySymbol_{#1}}
}
\newcommand{\energyPF}{\energy[PF]}
\newcommand{\energyHelfr}{\energy[H]}
\newcommand{\energyGL}{{\energy[GL]}}
\newcommand{\energyKin}{{\energy[K]}}
\newcommand{\Lagrangian}{\mathcal{L}}
\newcommand{\DissPotential}{\mathcal{D}}

\newcommand{\speedRS}[2][]
{
  \ifthenelse{\equal{#2}{}}
  {S_{#2}}
  {S_{#2}^{#1}}
}
\newcommand{\Interpolant}{I_{\meshparam}}

\newcommand{\ContSymbol}{C}
\newcommand{\Cont}[1][]{
  \ifthenelse{\equal{#1}{}}
  {\ContSymbol^{0}}
  {\ContSymbol^{#1}}
}
\newcommand{\Cinf}[1][]{
  \ifthenelse{\equal{#1}{}}
  {\ContSymbol^{\infty}}
  {\ContSymbol^{\infty}(#1)}
}
\newcommand{\SobSymbol}{W}
\newcommand{\HilbSymbol}{H}
\newcommand{\Hilb}[1][]{
  \ifthenelse{\equal{#1}{}}
  {\HilbSymbol^{1}}
  {\SobSymbol^{1}_{#1}}
}
\newcommand{\Sob}[2][]{
  \ifthenelse{\equal{#1}{}}
  {\HilbSymbol^{#2}}
  {\SobSymbol^{#2,#1}}
}

\newcommand{\LspaceSymb}{L}
\newcommand{\Lspace}[1][]{
  \ifthenelse{\equal{#1}{}}
  {\LspaceSymb^{2}}
  {\LspaceSymb^{#1}}
}
\newcommand{\Linfty}{\LspaceSymb^{\infty}}

\newcommand{\TestSpSymbol}{{V}}
\newcommand{\TestSpace}[1][]{
  \ifthenelse{\equal{#1}{}}
  {\TestSpSymbol({\SurfDomain})}
  {\TestSpSymbol_{#1}({\SurfDomain})}
}

\newcommand{\TestSpaceEmbedded}[1][]{
  \ifthenelse{\equal{#1}{}}
  {\TestSpSymbol(\TriangH(\SurfDomain))}
  {\TestSpSymbol_{#1}(\TriangH(\SurfDomain))}
}
\newcommand{\TestSpaceIntrinsic}[1][]{
  \ifthenelse{\equal{#1}{}}
  {\TestSpSymbol(\Triang(\SurfDomain))}
  {\TestSpSymbol_{#1}(\Triang(\SurfDomain))}
}
\newcommand{\TestSpaceChart}[1][]{
  \ifthenelse{\equal{#1}{}}
  {\TestSpSymbol(\Triang(\SubsetU))}
  {\TestSpSymbol_{#1}(\Triang(\SubsetU))}
}
\newcommand{\TestSpaceCell}[1][]{
  \ifthenelse{\equal{#1}{}}
  {\TestSpSymbol_{\meshparam}(\Cell)}
  {\TestSpSymbol_{\meshparam}(\Cell_{#1})}
}

\newcommand{\TestSpaceVol}{\TestSpace[\meshparam]}
\newcommand{\TestSpaceEmbeddedApprox}{\TestSpaceEmbedded[\meshparam]}
\newcommand{\TestSpaceChartApprox}{\TestSpaceChart[\meshparam]}

\newcommand{\TestSpaceApprox}[1][]{
  \ifthenelse{\equal{#1}{}}
  {\TestSpSymbol_{\meshparam}}
  {\TestSpSymbol_{\meshparam}(#1)}
}
\newcommand{\TestSpaceVec}[1][]{
  \ifthenelse{\equal{#1}{}}
  {\mathbf{\TestSpSymbol}_{\meshparam}}
  {\mathbf{\TestSpSymbol}_{\meshparam}(#1)}
}

\newcommand{\TestSpGammaSymbol}{{P}}
\newcommand{\TestSpaceGamma}[1][]{
  \ifthenelse{\equal{#1}{}}
  {\TestSpGammaSymbol_{\meshparam}}
  {\TestSpGammaSymbol_{\meshparam}(#1)}
}

\newcommand{\TestSpCHSymbol}{{W}}
\newcommand{\TestSpaceCH}[1][]{
  \ifthenelse{\equal{#1}{}}
  {\TestSpCHSymbol_{\meshparam}}
  {\TestSpCHSymbol_{\meshparam}(#1)}
}

\newcommand{\TestSpPressSymbol}{Q}
\newcommand{\TestSpacePress}[1][]{
  \ifthenelse{\equal{#1}{}}
  {\TestSpPressSymbol_{\meshparam}}
  {\TestSpPressSymbol_{\meshparam}(#1)}
}

\newcommand{\PolySymb}{\mathcal{P}}
\newcommand{\PC}[1]{\PolySymb{_{#1}}}
\newcommand{\PONE}{\PolySymb_{1}}
\newcommand{\PTWO}{\PolySymb_{2}}
\newcommand{\PZERO}{\PolySymb_{0}}


\newcommand{\GaussCurv}{\mathcal{K}}
\newcommand{\calH}{\mathcal{H}}
\newcommand{\meanCurv}[1][]
{
  \ifthenelse{\equal{#1}{}}
  {\calH}
  {\calH_{#1}}
}
\newcommand{\hmeanCurv}[1][]
{
  \ifthenelse{\equal{#1}{}}
  {\hat{\calH}}
  {\hat{\calH}_{#1}}
}
\newcommand{\meanCurvApprox}{\meanCurv[\meshparam]}
\newcommand{\shapeOp}{\mathcal{B}}
\newcommand{\AngleSymbol}{\theta}
\newcommand{\DevAngle}[1][]
{
  \ifthenelse{\equal{#1}{}}
  {\AngleSymbol}
  {\AngleSymbol_{\scriptscriptstyle{#1}}}
}
\newcommand{\relheightSymb}{\pi}
\newcommand{\relheight}[1][]
{
  \ifthenelse{\equal{#1}{}}
  {\relheightSymb_{\scriptscriptstyle{\SurfDomain}}}
  {\relheightSymb_{\scriptscriptstyle{#1}}}
}

\newcommand{\boundarySurf}{\SurfBnd}
\newcommand{\conormal}{\mathbf{\mu}}

\newcommand{\SolSymbol}{u}
\newcommand{\Sol}{\SolSymbol}
\newcommand{\AvgSol}{\bar{\SolSymbol}}
\newcommand{\forceSymbol}{f}
\newcommand{\force}{f}
\newcommand{\forceVecSymbol}{\mathbf{F}}
\newcommand{\forceVec}{\forceVecSymbol}
\newcommand{\AdvVel}{\Vector{w}}
\newcommand{\KinCoef}{\gamma}
\newcommand{\DiffCoef}{\epsilon}
\newcommand{\SolnCoef}{\DiffCoef}
\newcommand{\MassCoef}{c}
\newcommand{\StressTensComp}{\sigma}
\newcommand{\StressTens}{\boldsymbol{\StressTensComp}}
\newcommand{\DiffTens}{\mathbb{D}}
\newcommand{\DiffEig}[1][]
{ \ifthenelse{\equal{#1}{}}
  {d}
  {d_{#1}}
}
\newcommand{\DiffEigMax}{\DiffEig^*}
\newcommand{\DiffEigMin}{\DiffEig_*}

\newcommand{\BilinearStiffSymbol}{a}
\newcommand{\BilinearStiff}[3][]
{
  \ifthenelse{\equal{#1}{}}
  {\BilinearStiffSymbol(#2,#3)}
  {\BilinearStiffSymbol_{#1}(#2,#3)}
}
\newcommand{\BilinearAdvSymbol}{b}
\newcommand{\BilinearAdv}[3][]
{
  \ifthenelse{\equal{#1}{}}
  {\BilinearAdvSymbol(#2,#3)}
  {\BilinearAdvSymbol_{#1}(#2,#3)}
}
\newcommand{\BilinearMassSymbol}{m}
\newcommand{\BilinearMass}[3][]
{
  \ifthenelse{\equal{#1}{}}
  {\BilinearMassSymbol(#2,#3)}
  {\BilinearMassSymbol_{#1}(#2,#3)}
}
\newcommand{\BilinearReactSymbol}{c}
\newcommand{\BilinearReact}[3][]
{
  \ifthenelse{\equal{#1}{}}
  {\BilinearReactSymbol(#2,#3)}
  {\BilinearReactSymbol_{#1}(#2,#3)}
}
\newcommand{\MassForm}[2]{m(#1,#2)}
\newcommand{\Bilinear}[2]{a(#1,#2)}

\newcommand{\RhsSymbol}{F}
\newcommand{\Rhs}[1]{\RhsSymbol(#1)}

\newcommand{\TestSymbol}{v}
\newcommand{\Test}[1][]
{
  \ifthenelse{\equal{#1}{}}
  {\TestSymbol}  
  {\TestSymbol_{\scriptscriptstyle{#1}}}
}

\newcommand{\SolApprox}{\Sol_{\meshparam}}
\newcommand{\coeffSol}[1]{\Sol_{#1}}
\newcommand{\bcoeffSol}[1]{\bar{\Sol}_{#1}}
\newcommand{\AdvVelApprox}{\mathbf{w}_{\meshparam}}
\newcommand{\forceApprox}{\force_{\meshparam}}

\newcommand{\nodalBasis}[1]{\varphi_{\scriptscriptstyle{#1}}}
\newcommand{\exactBasis}[1]{\bar{\nodalBasis{#1}}}
\newcommand{\indecesBas}{\mathcal{I}}
\newcommand{\supportBas}{S}
\newcommand{\nNodes}[1][]
{
  \ifthenelse{\equal{#1}{}}
  {N^{\scriptscriptstyle{dof}}}
  {N^{\scriptscriptstyle{dof}}_{\scriptscriptstyle{#1}}}
}
\newcommand{\ndof}{3}
\newcommand{\nGPedge}{N^{\scriptscriptstyle{gp}}_{\scriptscriptstyle\Edge}}
\newcommand{\nGPcell}{N^{\scriptscriptstyle{gp}}_{\scriptscriptstyle\Cell}}

\newcommand{\TestApprox}[1][]
{
  \ifthenelse{\equal{#1}{}}
  {\TestSymbol_{\scriptscriptstyle{\meshparam}}}  
  {\TestSymbol_{\scriptscriptstyle{\meshparam,#1}}}
}
\newcommand{\coeffTest}[1]{\TestSymbol_{#1}}

\newcommand{\bTestApprox}[1][]
{
  \ifthenelse{\equal{#1}{}}
  {\bar{\TestSymbol}_{\scriptscriptstyle{\meshparam}}}  
  {\bar{\TestSymbol}_{\scriptscriptstyle{\meshparam,#1}}}
}

\newcommand{\TestZSymbol}{\mathbf{Z}}
\newcommand{\TestZ}[1][]
{
  \ifthenelse{\equal{#1}{}}
  {\TestZSymbol}  
  {\TestZSymbol_{\scriptscriptstyle{#1}}}
}

\newcommand{\TestZApprox}[1][]
{
  \ifthenelse{\equal{#1}{}}
  {\TestZSymbol_{\scriptscriptstyle{\meshparam}}}  
  {\TestZSymbol_{\scriptscriptstyle{\meshparam,#1}}}
}

\newcommand{\TestHSymbol}{{h}}
\newcommand{\TestH}[1][]
{
  \ifthenelse{\equal{#1}{}}
  {\TestHSymbol}  
  {\TestHSymbol_{\scriptscriptstyle{#1}}}
}

\newcommand{\TestHApprox}[1][]
{
  \ifthenelse{\equal{#1}{}}
  {\TestHSymbol_{\scriptscriptstyle{\meshparam}}}  
  {\TestHSymbol_{\scriptscriptstyle{\meshparam,#1}}}
}

\newcommand{\TestWSymbol}{w}
\newcommand{\TestW}[1][]
{
  \ifthenelse{\equal{#1}{}}
  {\TestWSymbol}  
  {\TestWSymbol_{\scriptscriptstyle{#1}}}
}

\newcommand{\TestWApprox}[1][]
{
  \ifthenelse{\equal{#1}{}}
  {\TestWSymbol_{\scriptscriptstyle{\meshparam}}}  
  {\TestWSymbol_{\scriptscriptstyle{\meshparam,#1}}}
}
\newcommand{\bTestWApprox}[1][]
{
  \ifthenelse{\equal{#1}{}}
  {\bar{\TestWSymbol}_{\scriptscriptstyle{\meshparam}}}  
  {\bar{\TestWSymbol}_{\scriptscriptstyle{\meshparam,#1}}}
}

\newcommand{\TestCHSymbol}{\psi}
\newcommand{\TestCHApprox}[1][]
{
  \ifthenelse{\equal{#1}{}}
  {\TestCHSymbol_{\scriptscriptstyle{\meshparam}}}  
  {\TestCHSymbol_{\scriptscriptstyle{\meshparam,#1}}}
}
\newcommand{\TestCH}[1][]
{
  \ifthenelse{\equal{#1}{}}
  {\TestCHSymbol}  
  {\TestCHSymbol_{\scriptscriptstyle{#1}}}
}
\newcommand{\TestCHmuSymbol}{\xi}
\newcommand{\TestCHmuApprox}[1][]
{
  \ifthenelse{\equal{#1}{}}
  {\TestCHmuSymbol_{\scriptscriptstyle{\meshparam}}}  
  {\TestCHmuSymbol_{\scriptscriptstyle{\meshparam,#1}}}
}

\newcommand{\TestVecSymbol}{\boldsymbol{v}}
\newcommand{\TestVecApprox}[1][]
{
  \ifthenelse{\equal{#1}{}}
  {\TestVecSymbol_{\scriptscriptstyle{\meshparam}}}  
  {\TestVecSymbol_{\scriptscriptstyle{\meshparam,#1}}}
}
\newcommand{\TestPressSymbol}{q}
\newcommand{\TestPress}[1][]
{
  \ifthenelse{\equal{#1}{}}
  {\TestPressSymbol}  
  {\TestPressSymbol_{\scriptscriptstyle{#1}}}
}
\newcommand{\TestPressApprox}[1][]
{
  \ifthenelse{\equal{#1}{}}
  {\TestPressSymbol_{\scriptscriptstyle{\meshparam}}}  
  {\TestPressSymbol_{\scriptscriptstyle{\meshparam,#1}}}
}

\newcommand{\BilinearStiffApprox}[2]{\BilinearStiffSymbol_{\meshparam}(#1,#2)}
\newcommand{\BilinearAdvApprox}[2]{\BilinearAdvSymbol_{\meshparam}(#1,#2)}
\newcommand{\BilinearReactApprox}[2]{\BilinearReactSymbol_{\meshparam}(#1,#2)}
\newcommand{\BilinearMassApprox}[2]{\BilinearMassSymbol_{\meshparam}(#1,#2)}
\newcommand{\RhsApprox}[1]{\RhsSymbol_{\meshparam}(#1)}

\newcommand{\BilinearApprox}[2]{a_{\meshparam}(#1,#2)}
\newcommand{\tBilinearApprox}[2]{\tilde{a}_{\meshparam}(#1,#2)}
\newcommand{\MassFormApprox}[2]{m_{\meshparam}(#1,#2)}
\newcommand{\tMassFormApprox}[2]{\tilde{m}_{\meshparam}(#1,#2)}

\newcommand{\tBilinearStiffApprox}[2]
{\tilde{\BilinearStiffSymbol}_{\meshparam}(#1,#2)}
\newcommand{\tBilinearAdvApprox}[2]
{\tilde{\BilinearAdvSymbol}_{\meshparam}(#1,#2)}
\newcommand{\tBilinearMassApprox}[2]
{\tilde{\BilinearMassSymbol}_{\meshparam}(#1,#2)}

\newcommand{\vectorRhs}{\mathbf{b}}
\newcommand{\vectorSol}{\mathbf{\Sol}}
\newcommand{\matrixStiff}{\mathbf{A}}
\newcommand{\matrixAdv}{\mathbf{B}}
\newcommand{\matrixReact}{\mathbf{C}}
\newcommand{\matrixMass}{\mathbf{M}}
\newcommand{\vectorRhsDG}{\mathbf{R}}

\newcommand{\tvectorRhs}{\tilde{\vectorRhs}}
\newcommand{\tmatrixStiff}{\tilde{\matrixStiff}}
\newcommand{\tmatrixAdv}{\tilde{\matrixAdv}}
\newcommand{\tmatrixMass}{\tilde{\matrixMass}}
\newcommand{\tmatrixReact}{\tilde{\matrixReact}}

\newcommand{\Peclet}[1]{\operatorname{Pe}_{\scriptscriptstyle{#1}}}
\newcommand{\stabCoef}[1]{\delta_{\scriptscriptstyle{#1}}}
\newcommand{\deltaO}{\delta_0}
\newcommand{\deltaI}{\delta_1}

\newcommand{\stabBil}[2]
{\operatorname{s}_{\scriptscriptstyle{\meshparam}}(#1,#2)}
\newcommand{\stabRhs}[1]{\operatorname{s}_{\RhsSymbol_{\meshparam}}(#1)}

\newcommand{\Stab}{\stabBil}

\newcommand{\entropySymb}{E}
\newcommand{\entropy}{\entropySymb}
\newcommand{\entropyVec}{\mathbf{\entropySymb}}
\newcommand{\entrFluxSymb}{G}
\newcommand{\entrFlux}{\mathbf{\entrFluxSymb}}

\newcommand{\massLumped}[1]{m_{#1}}
\newcommand{\Cijcoeff}[1]{\mathbf{c}_{#1}}
\newcommand{\artViscSymbol}{d}
\newcommand{\artVisc}[1]{\artViscSymbol_{#1}}

\newcommand{\viscositySymb}{\eta}
\newcommand{\viscosity}[1][]
{
  \ifthenelse{\equal{#1}{}}
  {\viscositySymb}
  {\viscositySymb_{\scriptscriptstyle{#1}}}
}

\newcommand{\ResidualSymbol}{R}
\newcommand{\Residual}[1][]
{
  \ifthenelse{\equal{#1}{}}
  {\ResidualSymbol}
  {\ResidualSymbol_{#1}}
}

\newcommand{\CPoincare}{C_{_{\SurfDomain}}}
\newcommand{\Ricci}{\operatorname{Ric}}
\newcommand{\Riccibound}{R}
\newcommand{\curvature}[1][]
{
  \ifthenelse{\equal{#1}{}}
  {\kappa}
  {\kappa_{#1}}
}
\newcommand{\Eig}{\lambda_1}

\newcommand{\QuadRule}[2][]
{
  \ifthenelse{\equal{#1}{}}
  {Q(#2)}
  {Q_{#1}(#2)}
}

\newcommand{\strainTens}{\mathbf{d}}

\newcommand{\mfrak}{\mathfrak{m}}
\newcommand{\wrt}{w.\,r.\,t.}
\newcommand{\st}{s.\,t.}
\newcommand{\ie}{i.\,e.}
\newcommand{\ia}{i.\,a.}
\newcommand{\oeda}{w.\,l.\,o.\,g.}
\newcommand{\resp}{resp.}
\newcommand{\cf}{cf.}

\newcommand{\formComma}{\,\text{,}}
\newcommand{\formPeriod}{\,\text{.}}

\newcommand{\xb}{\boldsymbol{x}}

\newcommand{\wb}{\vectvelV}
\newcommand{\ub}{\vectvel}
\newcommand{\vb}{\boldsymbol{\upsilon}}
\section{Introduction}
\label{sec:introduction}

Fluid deformable surfaces are ubiquitous in cell and tissue biology. Plasma membranes, the actomyosin cortical layer, or epithelial cell sheets, all share similar properties. They are thin fluidic sheets of soft materials exhibiting a solid–fluid duality. While they store elastic energy when stretched or bent, as solid shells, under in-plane shear, they flow as viscous two-dimensional fluids. This duality has several consequences: it establishes a tight interplay between tangential flow and surface deformation. In the presence of curvature, any shape change is accompanied by tangential flow and, vice versa, the surface deforms due to tangential flow \cite{ArroyoDesimone2009,Torres_Arroyo_2019,voigt2019fluid,al2021active,Krause_2023}. Of particular interest within morphogenesis and development is the inclusion of activity, leading to a variety of interesting morphodynamical phenomena and instabilities \cite{salbreux2017mechanics,al2021active,alIzzi}. However, due to computational challenges in dealing with these models, previous attempts only derive the equations or only address simplified cases. We here overcome these limitations and numerically consider models for fluid deformable surfaces under active geometric forces.

\begin{figure*}
    (a) \hspace*{7.3cm} (b) \hspace*{9.3cm} \\ \vspace*{-0.6cm}
    \includegraphics[width=0.44\linewidth]{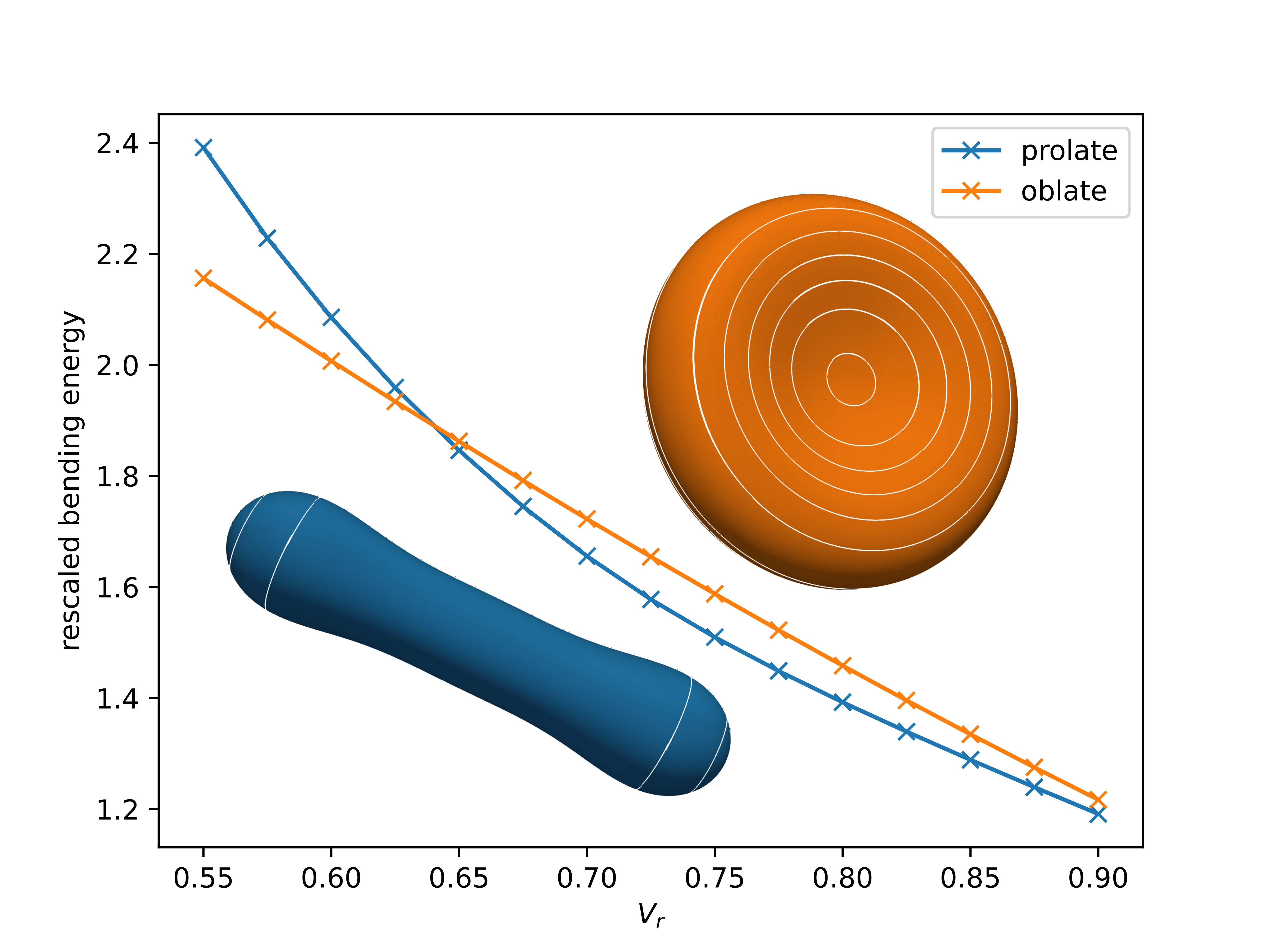}
    \includegraphics[width=0.55\linewidth]{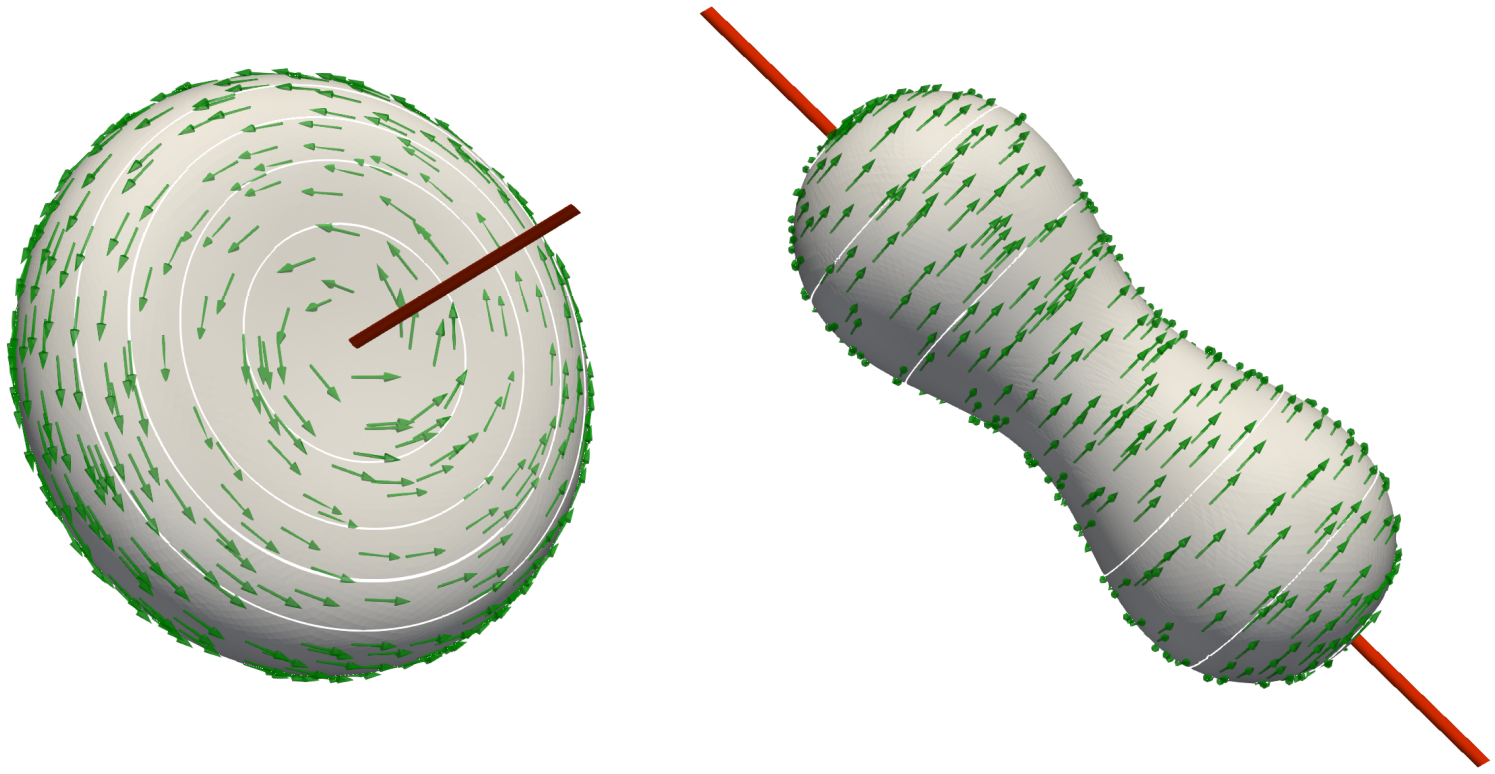}
  \caption{(a) Rescaled bending energy and examples of equilibrium shapes as a function of the reduced volume $V_r$ for $\Be = 1$. The shown shapes correspond to $V_r = 0.625$. Within the considered parameter space there are two stable branches of local minima. Their slopes and the transition between prolate and oblate shapes match the results in \cite{Seifert_AP_1997}. (b) Examples for axisymmetric equilibrium configurations for oblate- and prolate-like shapes with $V_r = 0.75$ and an angular velocity of $\omega = 3.864$ and $\omega = 2.324$, respectively. The axis denotes the rotational axis. The results correspond to the once reported in \cite{OlshanskiiEquilibriumStates,NestlerStability}. The visualizations show the shape together with isolines of the mean curvature (white lines) and the tangential flow field $\ProjMat \vectvel$ indicated by green arrows, see Section \ref{ssec:21} for detailed definitions. The results are obtained by simplifications of the numerical approach described in Section \ref{sec:3}.}
  \label{fig:1}
\end{figure*}

To explore the effect of these active geometric forces we investigate the stability of shapes obtained with classical membrane models. In these models the preferred shapes arise from a competition between curvature energy resulting from the bending elasticity of the membrane and geometrical constraints
such as fixed surface area and fixed enclosed volume \cite{Seifert_AP_1997}. These shapes of lowest energy can be classified in phase diagrams, see e.g. Figure \ref{fig:1} (a), which shows a turnover between prolate and oblate shapes. While such static aspects have been the focus of research for many years, within morphogenesis and development, dynamic aspects of shape transformations become crucial. Addressing the dynamics requires to also consider surface viscosity \cite{mongera2018fluid,PhysRevE.101.052401}, which leads to models for fluid deformable surfaces. They couple inextensible surface hydrodynamics with bending elasticity. In \cite{reuther2020numerical,Krause_2023} it is computationally shown that surface hydrodynmics significantly influences shape evolution, enhancing the evolution towards the equilibrium shapes and allowing to escape local minimum configurations. Within an axisymmetric setting also new equilibrium configurations, resulting from a balance of the normal components of the centrifugal forces and the bending forces, have been explored, see \cite{OlshanskiiEquilibriumStates}. Figure \ref{fig:1} (b) shows two examples of such configurations. However, the stationary flows in these configurations are surface Killing fields,  which only exist for axisymmetric surfaces. They can be interpreted as rigid body rotations around the symmetry axis and give rise to the mentioned centrifugal forces. The question arises if these shapes are also stable if the assumption on axisymmetry is dropped. While indeed such shapes can be reached also in the general setting (if started from an asisymmetric setting) \cite{reuther2020numerical,NestlerStability}, these shapes are not stable. Any shape change or perturbation of the velocity field will lead to dissipation and therefore destroy the force balance, which is explored numerically in \cite{NestlerStability}. Based on these results it has been postulated that due to thermal or numerical noise the considered states will relax towards the classical equilibrium shapes of \cite{Seifert_AP_1997} with the lateral motion of the membrane completely ceased.

We here extend these models for fluid deformable surface and incorporate active forces $\bm{F}_{a}$, see Eqs. \eqref{eq:fds1}-\eqref{eq:acf}. The simplest approach to incorporate activity is an active isotropic force $\bm{F}_{a} = \H \normalvec$, with $\H$ the mean curvature and $\normalvec$ the surface normal, see Section \ref{ssec:21} for detailed definitions. This force is tied to surface tension, influencing the energy reaction to changes in area \cite{Capovilla_2002}. Within various approaches this has been used to model mechanical feedback by considering the strength of the force as a function of a stress regulator molecule, see e.g. \cite{bois2011pattern,reymann2016cortical,mietke2019minimal,wittwer2023computational} for one-dimensional or axisymmetric settings. However, these approaches consider a compressible surface. Under inextensibility this active isotropic force has no effect on the solution. We therefore follow a different approach and consider the active geometric forces introduced in \cite{alIzzi}. They read $\bm{F}_{a} = \nabla_S \H + (\H^2 - 4\K) \normalvec$ for even forces and $\bm{F}_{a} = \mathbf{E} \cdot \nabla_S \H$ for odd forces, with $\K$ the Gaussian curvature and $\mathbf{E}$ the Levi-Civita tensor, see Section \ref{ssec:21} for detailed definitions.
These forces respond to mean curvature gradients. The even force acts like an active Marangoni-like stress \cite{salbreux2017mechanics} and the odd force acts in directions of constant mean curvature. The resulting odd stresses represent a novel form of odd elasticity characterized by normal deformations giving rise to asymmetric tangential forces. For special geometries the impact of these active geometric forces has been analysed \cite{alIzzi}. For tubes it could be shown that these forces can amplify perturbations of the membrane shape and give rise to shape instabilities. It is even speculated that such shape instabilities are the driving force behind various morphogenetic processes, e.g. the Drosophila hindgut, where an epithelial tissue tubule undergoes a chiral twisting motion to form a characteristic “question mark” shape \cite{doi:10.1126/science.1200940,10.7554/eLife.32506} or C. elegan zygotes, where chiral flows in the actomyosin cortical layer form a pseudocleavage furrow which drives axis convergence and initiate cell division \cite{10.7554/eLife.04165,bhatnagar2023axis}. We here explore these shape instabilities numerically for closed surfaces. Our starting configurations are the lowest energy shapes depicted in Figure \ref{fig:1} (a) with additional even and odd active geometric forces. Modifying the bending rigidity we explore the dynamics of these systems and categorize the resulting shapes and surface flow fields.

The paper is structured as follows: In Section \ref{sec:2} we provide the necessary notion and introduce the model, in Section \ref{sec:3} we describe the numerical algorithm, in Section \ref{sec:4} we demonstrate and discuss the simulation results with respect to shape instabilities, the resulting dynamic shape evolution and the emerging shapes and flow characteristics, and in Section \ref{sec:5} we discuss the results and relate them to processes in morphogenesis and development.

\section{Modeling}
\label{sec:2}

\subsection{Notation}
\label{ssec:21}

We follow the same notation as in \cite{Bachini2023} which is here repeated for convenience. We consider a time dependent smooth and oriented surface $\SurfDomain = \SurfDomain(t)$ without boundary, embedded in $\REAL^3$ and given via a parametrization $\param$. The enclosed volume is denoted by $\Omega = \Omega(t)$. We define the reduced volume as the scaled quotient of the embeded volume and the surface area $V_r = 6 \sqrt{\pi} |\Omega| / |\SurfDomain|^{3/2}$, see \cite{Seifert_AP_1997}. We denote by $\normalvec$ the outward pointing surface normal, the surface projection is $\ProjMat=\IDMat-\normalvec\otimes\normalvec$, with $\IDMat$ the identity matrix, the shape operator is $\shapeOp= -\GradP\normalvec$, the mean curvature $\meanCurv= \operatorname{tr}\shapeOp$, and the Gaussian curvature $\K = \frac{1}{2}\left(\H^2-\|\B\|^2\right)$. We consider time-dependent Euclidean-based $ n $-tensor fields in $T^n\REAL^3\vert_{\SurfDomain}$. We call $T^0\REAL^3\vert_{\SurfDomain} = T^0\SurfDomain$ the space of scalar fields, $T^1\REAL^3\vert_{\SurfDomain}=T\REAL^3\vert_{\SurfDomain}$ the space of vector fields, and $T^2\REAL^3\vert_{\SurfDomain} $ the space of 2-tensor fields.
Important subtensor fields are tangential n-tensor fields in $T^n\SurfDomain \le T^n\REAL^3\vert_{\SurfDomain}$.
Let $p \in T^0\SurfDomain$ (surface pressure) be a continuously differentiable scalar field, $\vectvel \in T\REAL^3\vert_{\SurfDomain}$ (surface velocity) a continuously differentiable $\REAL^3$-vector field, and $\StressTens \in T^2\REAL^3\vert_{\SurfDomain}$ (surface rate of deformation tensor) a continuously differentiable $\REAL^{3\times3}$-tensor field defined on $\SurfDomain$. We define the different surface gradients by $\GradP p = \ProjMat\nabla p^e$, $\GradP\vectvel = \ProjMat\nabla\vectvel^e\ProjMat$ and $\GradC\StressTens = \nabla\StressTens^e \ProjMat$,
where $p^e$, $\vectvel^e$ and $\StressTens^e$ are arbitrary smooth extensions of $p$, $\vectvel$ and $\StressTens$ in the normal direction and $\nabla$ is the gradient of the embedding space $\REAL^3$. The corresponding divergence operators for a vector field $\vectvel$ and a tensor field $\StressTens$ are $\DivP\vectvel = \operatorname{tr}(\GradP\vectvel)$ and $\DivC(\StressTens\ProjMat) = \operatorname{tr}\GradC(\StressTens\ProjMat)$, where $\operatorname{tr}$ is the trace operator. The relations to the covariant derivatives $\GradSurf$ and the covariant divergence $\divS$ on $\SurfDomain$, with $\Delta_{\S} = \divS \cdot \GradSurf$ the Laplace-Beltrami operator, read $\GradP p=\GradSurf p$ and
${\DivP\vectvel = \divS(\ProjMat\vectvel)-(\vectvel\cdot\normalvec)\meanCurv}$, respectively. Following the notation in \cite{alIzzi} we make use of the Levi-Civita tensor $\mathbf{E}$, whose action on a tangent vector $\vec{t} \in T\SurfDomain$ is given by $\mathbf{E}\cdot \Vec{t} = - \normalvec \times \Vec{t}$.

\subsection{Governing equations}

Within the Stokes limit the model for fluid deformable surfaces with conserved enclosed volume, friction and active geometric forces reads
\begin{align} \label{eq:fds1}
  0 &= -\GradP p-p\H\bm{\nu} + \frac{2}{\Reyn} \div_C\bm{\sigma} - L R \bm{u} + \bm{b}_N - \lambda\bm{\nu} \nonumber \\
  & \quad \,+ \Ac \bm{F}_{a} \\
   0 &= \DivP \bm{u} \label{eq:fds2} \\
   0 &= \int_\S \bm{u}\cdot\bm{\nu}\,d\S
\end{align}
where $p$ is the surface pressure, $\vectvel$ the surface velocity and
\begin{align} \label{eq:fds4}
  \StressTens(\vectvel) &= \frac{1}{2} (\GradP \vectvel + (\GradP \vectvel)^T) \\
  \label{eq:fds5}
  \bm{b}_N &= - \frac{1}{\Be} \left( \Delta_\S \H + \H \left(\|\B\|^2-\frac{1}{2} \H^2\right) \right)\normalvec
\end{align}
the rate of deformation tensor and the bending force, respectively. The active geometric forces are defined by
\begin{equation}\label{eq:acf}
  \bm{F}_{a} =
  \begin{cases}
    \GradP \H + (\H^2 - 4\K) \normalvec & \mbox{(even)}\\
    \mathbf{E} \cdot \GradP \H & \mbox{(odd)}
  \end{cases}\,.
\end{equation}
The parameters are $\Reyn$ Reynolds number, $\Be$ bending capillary number, $\Ac$ activity number, $L R$ friction coefficient and $\lambda$ is a Lagrange multiplier to ensure a constant enclosed volume $|\Omega|$. 

Eqs. \eqref{eq:fds1} - \eqref{eq:fds5} with $\Ac = 0$ have been introduced and numerically solved in~\cite{Torres_Arroyo_2019,Krause_2023,NestlerStability}. For further numerical and analytical approaches under additional symmetry assumptions we refer to \cite{ArroyoDesimone2009,BORJADAROCHA2022104876,OlshanskiiEquilibriumStates}. The active forces in Eq. \eqref{eq:acf} have been proposed in \cite{alIzzi}. The even form effectively only contributes a normal force. This becomes evident by identifying $\tilde{p} =  p - \Ac \H $ as a generalized surface pressure, which does not impact the velocity field $\vectvel$. The odd form contributes a tangential force in response to gradients of mean curvature but acting at right angles to them. This can be interpreted as a geometric counterpart of odd elasticity \cite{annurev:/content/journals/10.1146/annurev-conmatphys-040821-125506}. Both forms are the lowest order contributions considered in \cite{alIzzi}. While these active geometric forces have been phenomenologically proposed and motivated by active gel theories \cite{julicher2018hydrodynamic} with a polarisation in normal direction in \cite{alIzzi}, they can also be derived from general active nematic forces in surface active nematodynamics \cite{nitschke2019hydrodynamic,nestler2022active,nitschke2024active}. In this context they are associated with a tangential nematic field aligned along the principal curvature directions. Such alignment results from the geometric coupling to extrinsic curvature contribution in surface Oseen-Frank or Landau-de Gennes energies \cite{napoli2012surface,golovaty2017dimension,nestler2018orientational,Nitschke_2018,Nitschke_2020}. Coupling terms between the director field and the tangential flow field lead to alignment of both and therefore active flows in principle curvature directions \cite{nitschke2024active}. \\

\section{Numerical approach}
\label{sec:3}

We follow the numerical approaches of \cite{Krause_2023,Bachini2023}, who consider Eqs. \eqref{eq:fds1} - \eqref{eq:fds5} with $\Ac = 0$ and $LR = 0$ and instead of the Stokes limit the full surface Navier-Stokes equations. The reduction from the Navier-Stokes to the Stokes equations is straightforward and the additional friction and the active forces can be incorporated easily as well. The approach considers a surface finite element method (SFEM) \cite{dziuk2013finite,nestler2019finite} in space and a semi-implicit finite difference method in time within an Arbitrary Lagrangian-Eulerian (ALE) approach \cite{ElliottStyles_ALE_2012}, Lagrangian in normal direction and Eulerian in tangential direction.

We combine Eqs. \eqref{eq:fds1} - \eqref{eq:fds5} with a mesh redistribution approach, see \cite{Barrett_ParametricFE2008}. These are equations for the parametrization
\begin{align}
        \DerT \param \cdot \normalvec &= \vectvel\cdot\normalvec\\
        \meanCurv \normalvec &= \Delta_C \param \,,
\end{align}
which provide the relation between the surface fluid velocity $\vectvel$ and the parameterization $\param$ and additionally provide an implicit representation of the mean curvature $\H$. We consider a discrete $k$-th order approximation $\SurfDomain_h^k$ of $\SurfDomain$, with $h$ the size of the mesh elements, i.e. the longest edge of the mesh. We consider each geometrical quantity like the normal vector $\normalvec_h$, the shape operator $\shapeOp_h$, the Gaussian curvature $\K_h$, and the inner products $(\cdot , \cdot)_h$ with respect the $\SurfDomain_h^k$. In the following we will drop the index $k$ and only write $\SurfDomain_h$. We define the discrete function spaces for scalar functions by $V_{k}(\SurfDomain_h)=\{ \psi \in C^0(\SurfDomain_h) \vert \psi\vert_{\Cell}\in\mathcal{P}_{k}({\cal{T}})\}$ and for vector fields by $\boldsymbol{V}_{k}(\SurfDomain[\meshparam])=[V_{k}(\SurfDomain[\meshparam])]^3$. Within these definitions $\cal{T}$ is the mesh element and $\mathcal{P}_{k}$ are the polynomials of order $k$. We consider $\vectvel_h,\param\in\boldsymbol{V}_3(\SurfDomain_h)$,  $\meanCurv_h,\in V_3(\SurfDomain_h)$, and $p_h\in V_2(\SurfDomain_h)$, which leads to an isogeometric setting for the velocity and a $\mathcal{P}_{3}-\mathcal{P}_{2}$ Taylor-Hood element for the surface Stokes-like equations. We discretize in time using a semi-implicit time stepping scheme with constant time stepping with step size $\tau$. In each time step $t^n$ we solve the surface Stokes-like equations and the mesh redistribution together. We define a discrete surface update variable $\update^{n}=\param^{n}-\param^{n-1}$, which is considered as unknown instead of the surface parametrization $\param^{n}$. Treating all geometric terms with respect to $\param^{n-1}$ leads to a linear system, which reads: \\

\begin{widetext}
\noindent
Find
$(\vectvelApprox^n,\pressApprox^n,\meanCurv_h^n,\update^n)\in[\boldsymbol{V}_3\times V_2\times V_3 \times \boldsymbol{V}_3](\SurfDomain_h^{n-1})$ such that:
\begin{align*}
  0 &= \InnerApprox{\pressApprox^n}{\DivP\TestVecApprox}
   - \frac{2}{\Reynolds} \InnerApprox{
   \StressTens(\vectvelApprox^{n})}{\GradP\TestVecApprox} - LR \InnerApprox{\vectvel_h^n}{\TestVecApprox}  + \frac{1}{\Be} \InnerApprox{ \GradSurf \meanCurv_h^n }{\GradSurf(\TestVecApprox\cdot\normalvec_h^{n-1})} +  \frac{1}{\Be} \InnerApprox{\meanCurv_h^nB^{n-1} \normalvec_h^{n-1}}{\TestVecApprox}
 \\
 &\quad + \lambda \InnerApprox{\normalvec_h^{n-1}}{\TestVecApprox} +  \Ac \InnerApprox{\bm{F}_{a,h}^{n,n-1}}{\TestVecApprox}\\
0 &=\InnerApprox{\DivP\vectvelApprox^{n}}{\TestPressApprox}\\
    0 &= \InnerApprox{\meanCurv_h^n\normalvec_h^{n-1}}{\TestZApprox}
    + \InnerApprox{\GradC \update^n}{\GradC \TestZApprox}
    + \InnerApprox{\GradC \MapU^{n-1}}{\GradC \TestZApprox}
\end{align*}
for all $(\TestVecApprox,\TestPressApprox,\TestHApprox,\TestZApprox)\in[\boldsymbol{V}_3\times V_2\times V_3\times\boldsymbol{V}_3](\SurfDomain_h^{n-1})$, where
$B^{n-1}=\Vert\shapeOp_h^{n-1}\Vert^2-\frac{1}{2} \left(\tr \shapeOp_h^{n-1}\right)^2$ and $\bm{F}_{a,h}^{n,n-1} = \GradSurf \H_h^n + (\H_h^n \H_h^{n-1} - 4 \K_h^{n-1}) \normalvec_h^{n-1}$ for the even case and $\bm{F}_{a,h}^{n,n-1} = \mathbf{E}_h^{n-1} \cdot \nabla_S \H_h^{n-1}$ for the odd case. \\
\end{widetext}

In the above formulation we used the identity $\InnerApprox{-\GradSurf p_h^n - p_h^n \meanCurv_h^n \normalvec_h}{\TestVecApprox} = \InnerApprox{p_h^n}{\DivP \TestVecApprox}$. Note that the Lagrange multiplier $\lambda$ is still unknown, which leads to an underdetermined system. To resolve that problem, we follow the approach introduced in \cite{Krause_2023}.  In order to fulfill the volume constraint, $\lambda$ has to be chosen such that
\begin{align}
	\Phi(\lambda)\coloneqq \int_{\S^{n-1}_h} \mathbf{u}^{n}_h(\lambda)\cdot \bm{\nu}_h^{n-1}\,d\S = 0.
\end{align}
We consider $\Phi(\lambda) = 0$ as an equation in $\lambda$ and apply a Newton iteration $\lambda^{j+1} = \lambda^{j} - \Phi(\lambda^{j}) / \Phi'(\lambda^{j})$. Each iteration thus requires to solve the surface Stokes-like equations. After convergence is achieved the last step requires to compute the new surface $\S_h^n$ by updating the parametrization $\param^{n} =  \param^{n-1} + \update^{n}$, lifting the solutions $\vectvel_h^n$, $\press_h^n$ and $\meanCurv_h^n$ to the new surface and to compute the remaining geometric quantities $\normalvec_h^{n}$, $\shapeOp_h^n$ and $\K_h^n$ for the new surface.

\begin{table}
  \begin{tabular}{c|c|c}
$h$&$\mathrm{eoc}(|F_{bend,h} -F_{bend,h_{ref}}|)$&$\mathrm{eoc}(\norm{\DivP \vectvel_h}_{L^2})$\\
\hline
$0.1854$&$5.99$&$11.57$\\
$0.119978$&$2.95$&$5.58$\\
$0.0927002$&$3.37$&$3.27$
  \end{tabular}
  \caption{Experimental order of convergence ($\mathrm{eoc}$) of errors considered for the evolution of a prolate under an odd force with parameters $\Ac = \Reynolds = 1, \Be = 100$ and timestepsize $\tau \sim h^3$. The errors are measured at time $T=2$, where a significant shape change has happened, but before the evolution determines in an attracting fixpoint. Note, that only $\norm{\DivP \vectvel_h}_{L^2}$ is an independent error measurement, while in $|F_{bend,h} -F_{bend,h_{ref}}|$ we consider the error with respect to a finer solution computed for $h_{ref} = 0.059989$. \label{tab:1}}
\end{table}

The semi-implicit or explicit treatment of $\bm{F}_{a,h}^{n,n-1}$ is a result of numerical experiments. For $\Ac = 0$ the approach corresponds to the one used in \cite{Krause_2023} for which presumably optimal order of convergence has been experimentally shown. As in \cite{Krause_2023} the discretization is realized within the finite element toolbox AMDiS \cite{vey2007amdis,witkowski2015software} using the \textsc{Dune}-CurvedGrid library \cite{praetorius2020dunecurvedgrid}. For $\Ac = 1$ we demonstrate (within the limits which are computationally achievable with reasonable effort) the same experimental order of convergence as reported in \cite{Krause_2023}. For the same combination $\vectvel_h,\param\in\boldsymbol{V}_3(\SurfDomain_h)$ and $p_h\in V_2(\SurfDomain_h)$ third order convergence for the inextensibility error $\norm{\divC \vectvel}_{L^\infty(L^2)}$ has been analytically shown to be optimal for Stokes flow on stationary surfaces \cite{hardering2023parametric,reusken2024analysis}. Theoretical results for Willmore flow without any constraints with the considered approach $\param\in\boldsymbol{V}_2(\SurfDomain_h)$ and $\meanCurv_h,\in V_2(\SurfDomain_h)$ indicate second order convergence for $|F_{bend,h} - F_{bend}|$ \cite{Dziuk_NM_2008}. The results in Table \ref{tab:1} show the experimental order of convergence ($\mathrm{eoc}$) of various quantities evaluated with respect to the analytical solution or a finer reference solution. While they cannot be used to claim optimal order of convergence, they at least demonstrate the validity of the numerical approach. The obtained orders of convergence for the finest resolution are similar to the once shown in \cite{Krause_2023} without active forces.

\section{Simulation Results}
\label{sec:4}

\subsection{Stability of minimizers}
\label{ssec:41}

\begin{figure}[h]
    \begin{flushleft}
    (a) \hspace*{4cm} (b)
    \end{flushleft} \vspace*{-0.3cm}
    \includegraphics[width=0.49\linewidth]{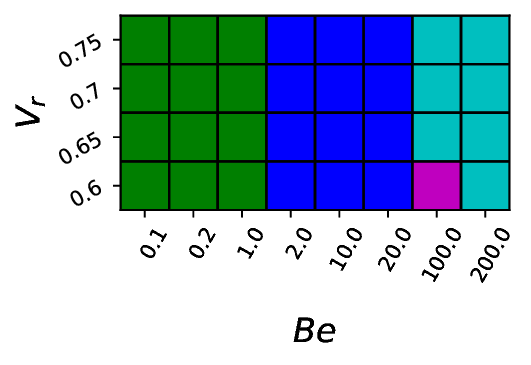}
    \includegraphics[width=0.49\linewidth]{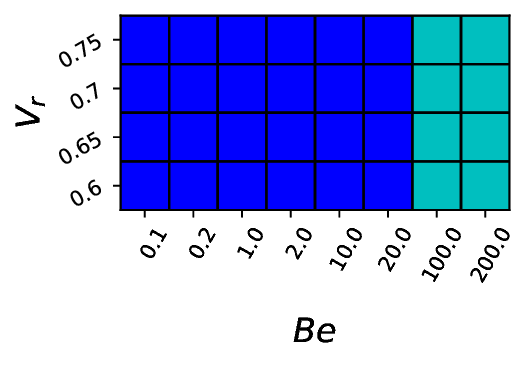} \\
    \vspace*{-0.5cm}
    \begin{flushleft}
    (c) \hspace*{4cm} (d)
    \end{flushleft} \vspace*{-0.3cm}
    \includegraphics[width=0.49\linewidth]{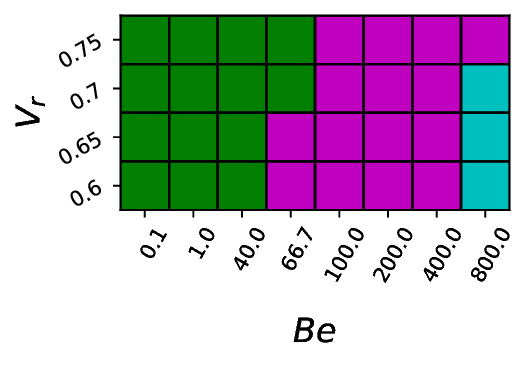}
    \includegraphics[width=0.49\linewidth]{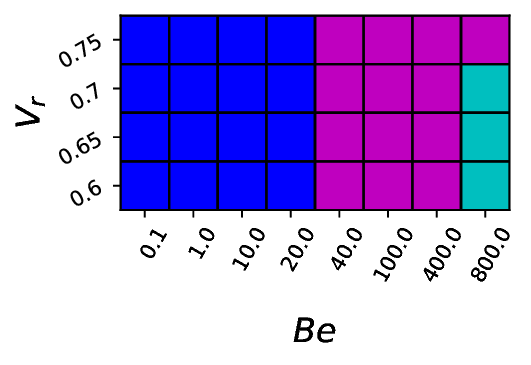}
  \caption{Stability diagram for $\Ac = 1$. (a) even forces and initial prolate shape. (B) even forces and initial oblate shape. (c) odd forces and initial prolate shape. (d) odd forces and initial oblate shape. The colors indicate the resulting shapes: green - prolate-like, blue - oblate-like, purple - different shape, turquoise - simulation crashed.}
  \label{fig:2}
\end{figure}

\begin{figure*}
    \centering
    \includegraphics[width=\linewidth]{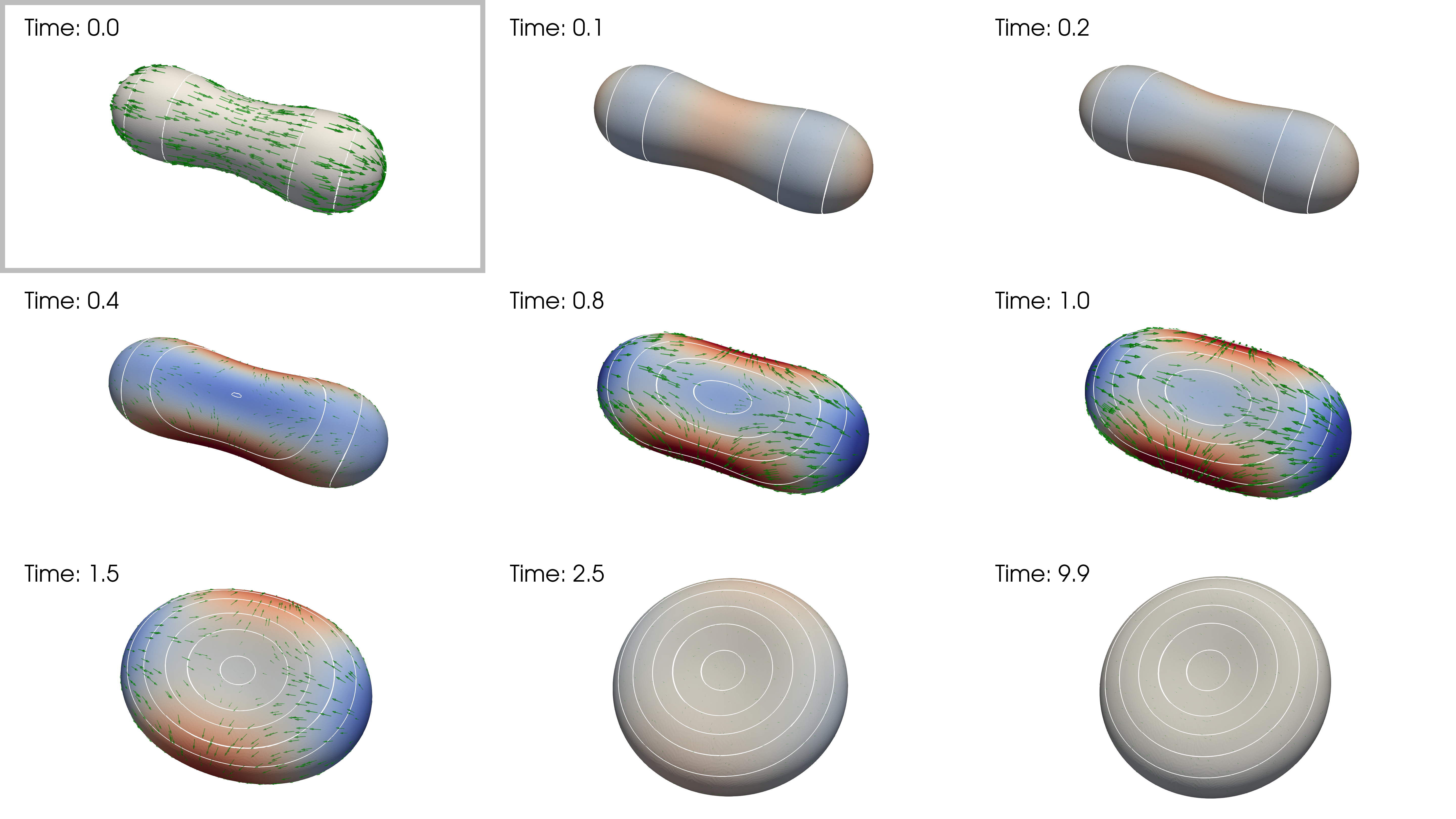}
    \caption{Shape evolution for $\Ac = 1$, $V_r = 0.75$ and an even force with $\Be = 2$ starting from a prolate shape. The visualizations show the shapes together with isolines of the mean curvature (white lines), the tangential flow field $\ProjMat \vectvel$ indicated by green arrows with the magnitude indicated by the length and the movement in normal direction indicated by blue (inwards) and red (outwards) at different time instances. The scaling is the same for all snapshots except for the initial condition. Here the magnitude of the tangential flow is enlarged. (Multimedia available online).}
    \label{fig:3}
\end{figure*}

We report on the stability of equilibrium shapes, as shown in Figure \ref{fig:1} (a). These shapes, which result as minimizers of the bending energy $F_{bend} = \frac{1}{2\Be} \int_{\SurfDomain} \H^2 \;d \SurfDomain$ under the constraints of fixed surface area $|\SurfDomain|$ and fixed enclosed volume $|\Omega|$, or if following the argumentation in \cite{NestlerStability} as equilibrium solutions of Eqs. \eqref{eq:fds1} - \eqref{eq:fds5} with $\Ac = 0$, serve as initial shapes. We consider even and odd active geometric forces with $\Ac = 1$. In Figure \ref{fig:2} the resulting stability diagrams are shown for various bending capillary numbers $\Be$ and reduced volumes $V_r$. The considered parameter range is restricted to $V_r \in [0.6,0.8]$. For smaller values potential self-intersections of the surface would cause difficulties. Preventing self-intersections would require modifications of the numerical algorithm. While various approaches have been proposed \cite{Yu:2021:RS,bartels2022simulating,Sassen:2024:RS}, within the context of fluid deformable surfaces we are not aware of a numerical realization. We therefore refrain from exploring smaller values. Larger values simply do not lead to any interesting shape evolution. The considered parameter range for $\Be$ spans the numerically accessible range. As bending is the dominating regularization in fluid deformable surface models, values above a upper threshold (corresponding to low bending rigidities) can lead to numerical instabilities. On the other hand, below a lower threshold the initial shapes are expected to be stable.

For even active geometric forces the results indicate almost no dependency on $V_r$. However, varying $\Be$ leads to different behaviours. Starting from a prolate shape, three regimes can be identified. For small values ($\Be \lesssim 1$) the initial shape is stable, it remains in a prolate-like configuration. For larger values ($1 \lesssim \Be \lesssim 20$) the initial shape is unstable and evolves to an oblate-like shape. An examples of the shape evolution within the unstable regime is shown in Figure \ref{fig:3}. For very large values ($\Be \gtrsim 100$) the active force overcomes bending, viscosity and friction, leading to disc-like shapes and rigid motions. Ultimatively, this behaviour causes strong mesh dilation and the numerical simulations tend to crash after some time. Starting from an oblate shape only two regimes can be identified. For values $\Be \lesssim 20$ the initial shape is stable and for larger values the numerical solution becomes unstable.

\begin{figure*}
    \centering
    \includegraphics[width=\linewidth]{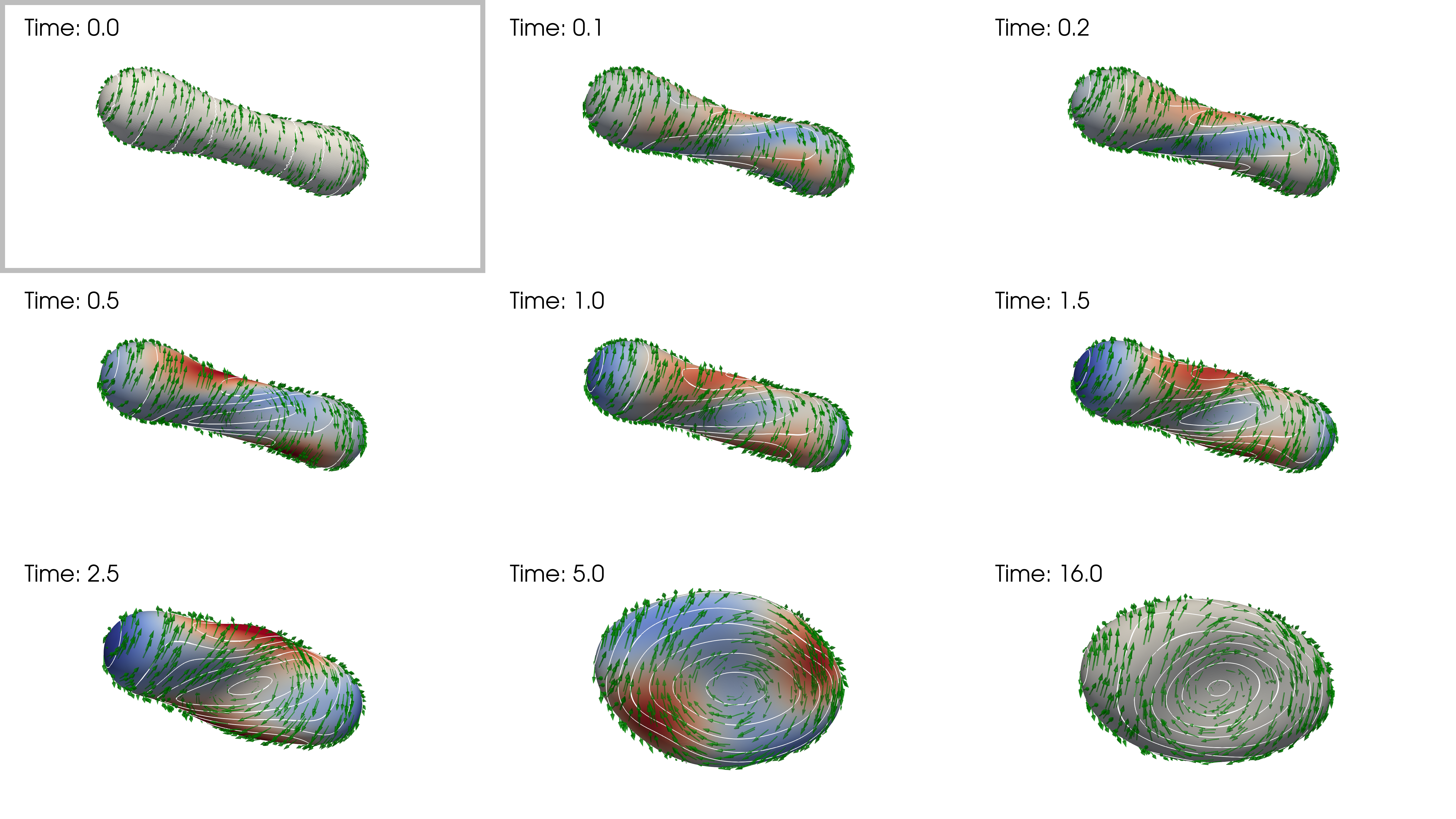}
    \caption{Shape evolution for $\Ac = 1$, $V_r = 0.65$ and an odd force with $\Be = 200$ starting from a prolate shape. The visualizations follow the same rules as in Figure \ref{fig:3}. (Multimedia available online).}
    \label{fig:4}
\end{figure*}
\begin{figure*}
    \centering
    \includegraphics[width=\linewidth]{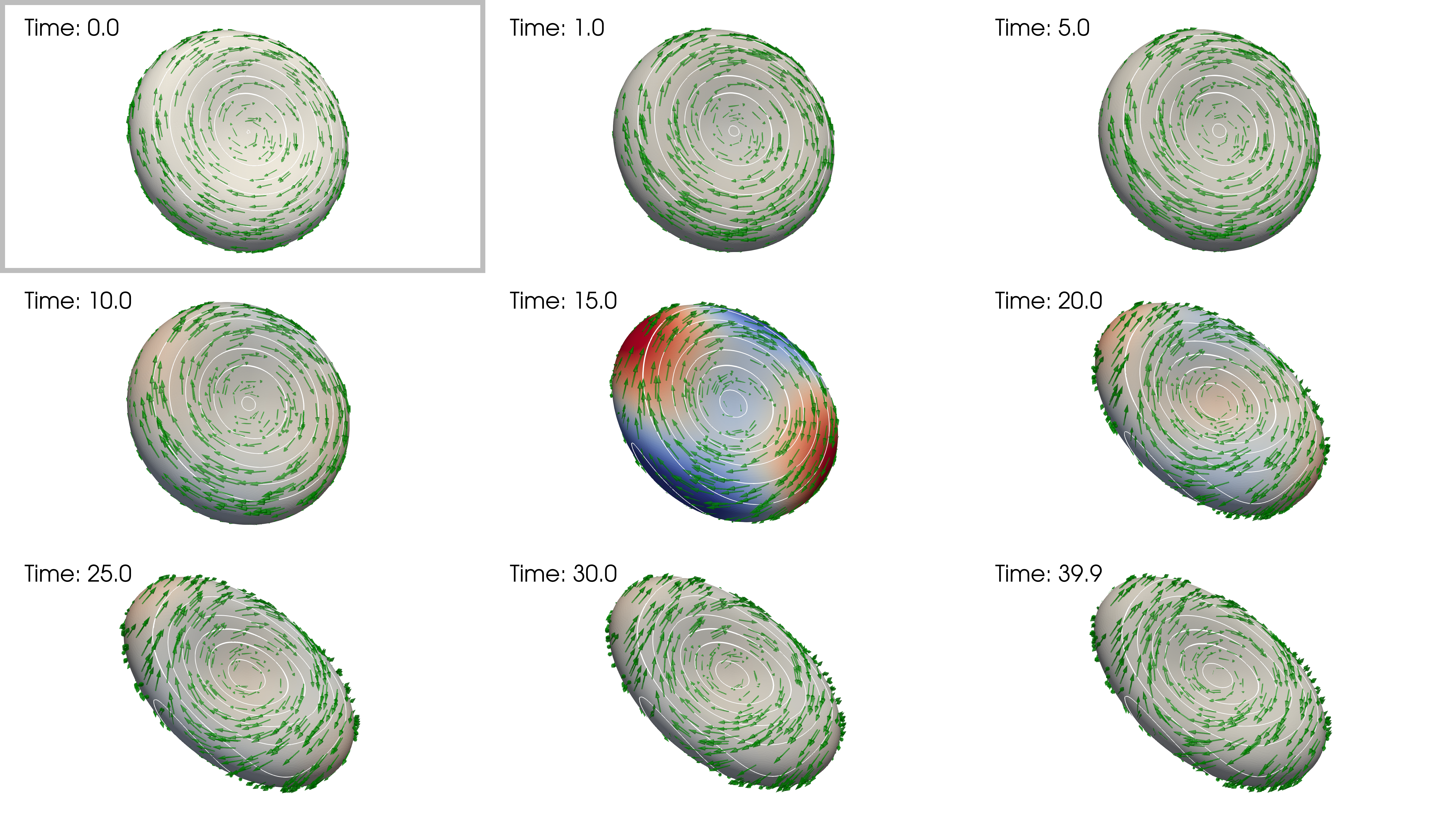}
    \caption{Shape evolution for $\Ac = 1$, $V_r = 0.65$ and an odd force with $\Be = 200$ starting from a oblate shape. The visualizations follow the same rules as in Figure \ref{fig:3}. (Multimedia available online).}
    \label{fig:5}
\end{figure*}

For odd active geometric forces the stability diagrams are more complex. Starting from a prolate shape we again have a regime for which the initial shape is stable. This happens for $\Be \lesssim 40$. The value slightly increases for increasing $V_r$. For larger bending capillary numbers the initial shape is unstable and evolves to a new shape. Only for very large values ($\Be \gtrsim 400$) the numerical solution becomes unstable, at least for low $V_r$. The qualitative behaviour is similar if we start from an oblate shape. For $\Be \lesssim 20$ the initial shape is stable and for larger bending capillary number the initial shape is unstable and evolves to a new shape. Again the numerical solution becomes unstable for very large values ($\Be \gtrsim 400$). Two examples of the shape evolution within the unstable regimes are shown in Figures \ref{fig:4} and \ref{fig:5}, starting from a prolate and an oblate initial shape, respectively.

Theoretical foundations for the observed mechanical instabilities only exist in simplified cases. In \cite{alIzzi} a stability analysis is considered for an infinite tube. For an odd active geometric force a self-reinforcing helical instability is found. Perturbations in shape induce chiral flows which enhance the perturbations. This is analysed with respect to the Scriven-Love number SL = $\Reynolds / \Be$ \cite{PhysRevE.101.052401}, which comprises viscous forces in the normal direction, which arise due to the coupling between in-plane viscous stresses and membrane curvature, to bending forces. Our simulation results complement those studies and provide realisations for closed systems.

In addition the simulation results also indicate a related mechanical instability for even active geometric forces. However, the qualitative differences between Figure \ref{fig:2} (a) and (b) indicate that oblate-like shapes are more stable than prolate-like shapes against even active geometric forces.

\subsection{Dynamic properties and emerging shapes}

Figure \ref{fig:2} not only addresses the stability of the initial configurations, but also the emerging shapes. We now aim to relate these emerging shapes to the dynamics they encountered before the shapes are reached. We therefore analyse the tangential flow fields for odd forces in more detail. In contrast with the equilibrium shapes in Figure \ref{fig:1} (b) with a constant angular velocity, we initially have chiral flows in Figures \ref{fig:4} and \ref{fig:5}. As mentioned in \cite{alIzzi}, the direction of induced tangential flows reverses between prolate and oblate shapes. The odd active geometric forces generate circulating flows that are right handed for a prolate shape and left handed for an oblate shape. As analysed in \cite{alIzzi} the position of maximum velocity is that of the highest gradient in mean curvature, which is closer to the poles in the case of a prolate shape and closer to the equator for an oblate shape. This behaviour is found in our simulations, see the initial configurations in Figures \ref{fig:4} and \ref{fig:5} respectively.

Due to the topology of the considered shapes the tangential flow field, as being non-zero, has to have defects, which can be identified as vortices and saddle-points in the vorticity field. Various numerical criteria exist to identify these defects in flat space \cite{Chakraborty,Kolar}. We here adapt the Q-criterion to identify vortices, with
$Q = \frac{1}{2}\left(\norm{\StressTens}^2 - \norm{\boldsymbol{\omega}}^2\right)$ and $\boldsymbol{\omega} = \frac{1}{2}(\GradP \vectvel - (\GradP \vectvel)^T)$ the vorticity tensor. We identify connected fluid regions where the vorticity magnitude prevails over the strain-rate magnitude $Q<-\epsilon$, with $\epsilon = 0.01$ experimentally chosen. Each of those regions is considered a vortex if the tangential velocity field $\ProjMat \vectvel$ vanishes up to a numerical tolerance. We count the number of vortices. The corresponding number of saddle-points follows from the topological constraint. This algorithm produces reasonably good results and only has problems if defects merge or split. This might be cured by further tuning the tolerances, but is not relevant for our purpose of classifying the dynamics.
\begin{figure}
    \includegraphics[width=\linewidth]{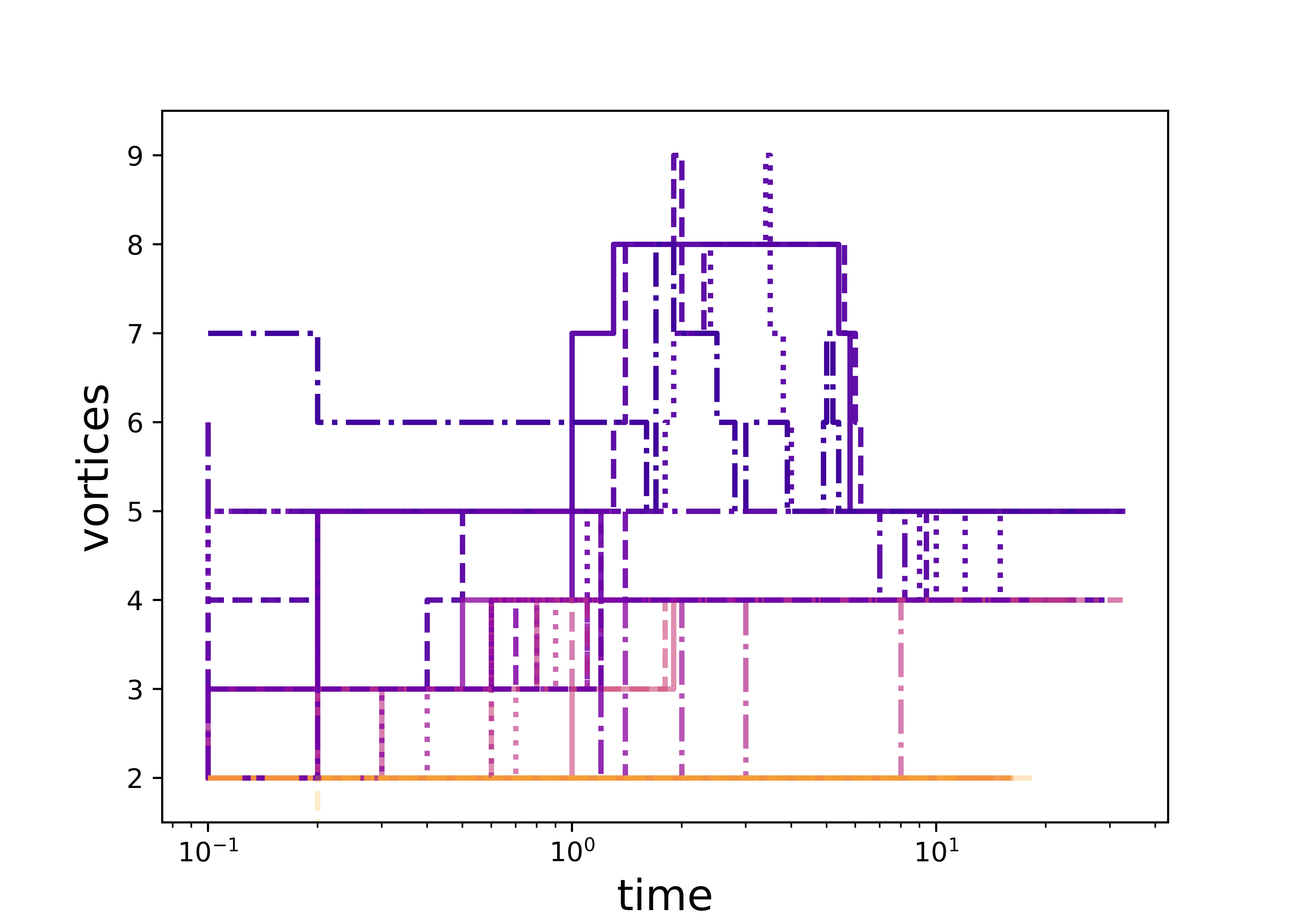}
    \hspace*{-0.7cm}
    \includegraphics[width=0.9\linewidth]{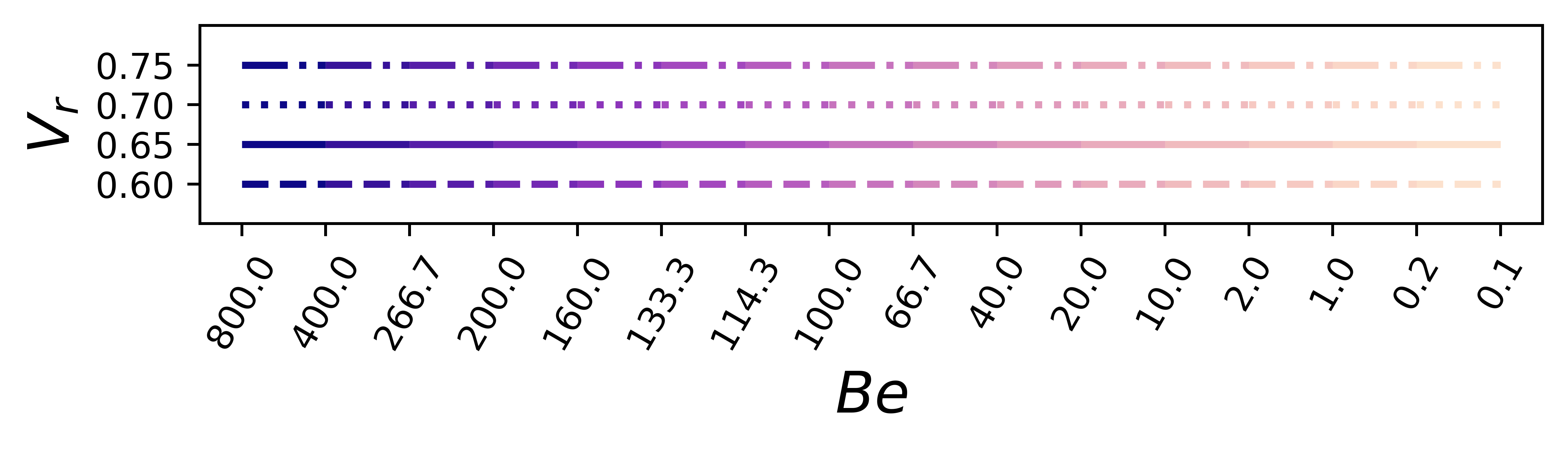}
    \caption[Vortex dynamics]{Creation and annihilation of vortices under odd force. Number of vortices over time for the same range of values for $\Be$ and $V_r$ as in Figure \ref{fig:2} (c), except for numerical instabilities. The results indicate the rich dynamics in the tangential flow field following the shape instabilities and the stability of additional vortices also in the reached equilibrium states. \label{fig:6}}
\end{figure}
In Figure \ref{fig:6} the number of vortices is plotted over time for the parameters considered in Figure \ref{fig:2} (c). The dynamics is manifold. Two counterrotating vortices are immediately formed. For the regime $\Be \lesssim 40$ the initial shape is stable and also the number of vortices does not change over time. For larger values however, these counterrotating flows cause spiraling bulges along the prolate, the number of which seems to be controlled by the ratio of bending and active forces. Even thought larger $\Be$ allow for more of those chiral lines, in the simulated regimes they merge to usually two or three spirals that actually cause outwards buckling. These shape changes create curvature gradients between the bulges and thus induce flow. This leads to the formation of additional vortices. Essentially for all cases with $\Be < 200$, this initial phase results in two spirals, which untwist the shape towards an elongated biconcave shape with a total of four vortices and two saddle points. An example of this type of evolution is depicted in Figure \ref{fig:4}. For larger values, the behaviour changes leading to three spirals.
\begin{figure*}
    \centering
    \includegraphics[width=\linewidth]{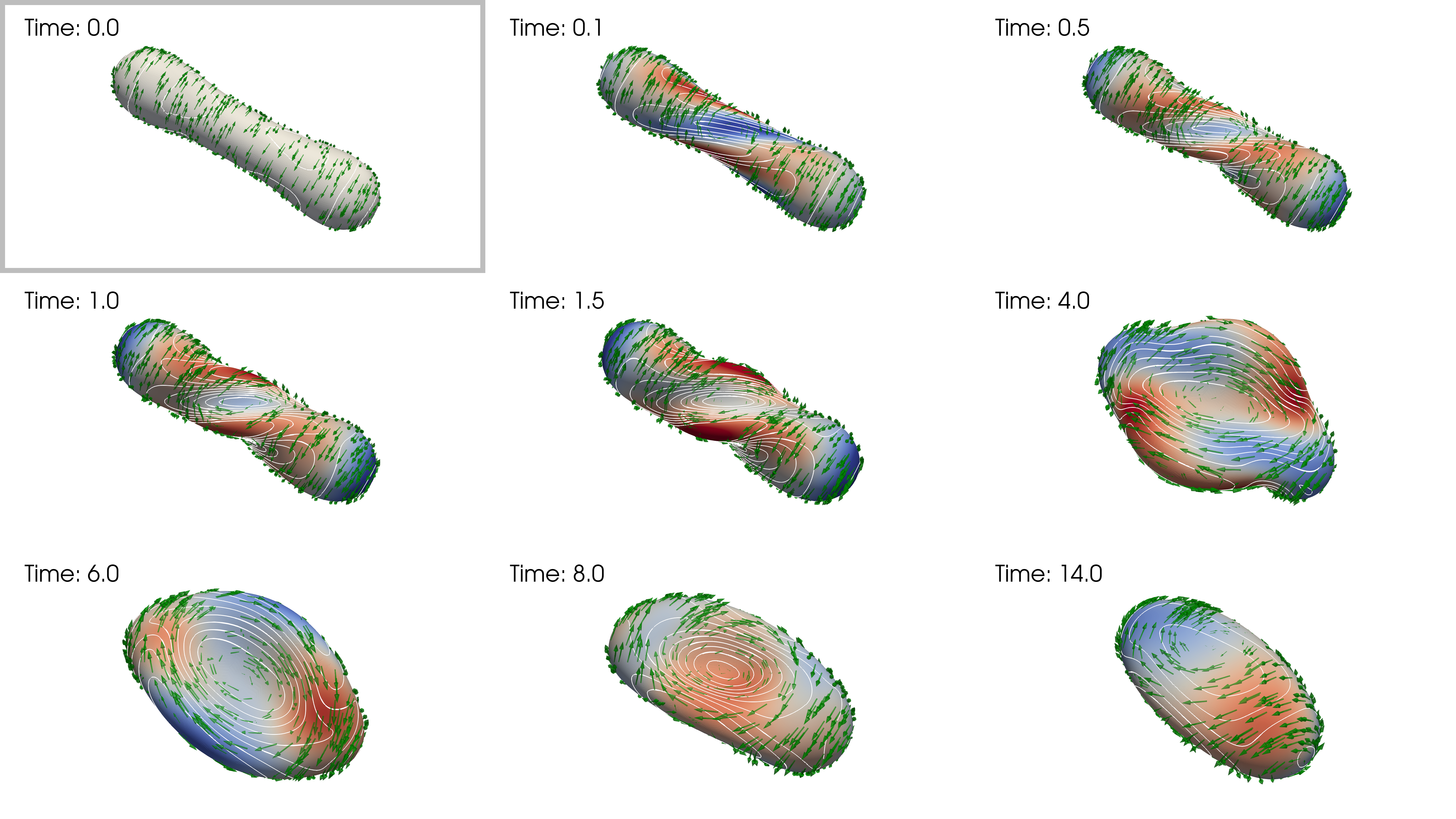}
    \caption{Shape evolution for $\Ac = 1$, $V_r = 0.65$ and an odd force with $\Be = 400$ starting from a prolate shape. The visualizations follow the same rules as in Figure \ref{fig:3}. The coloring indicating movement in normal direction also at the last time instance results from rigid body motion of this shape. (Multimedia available online) .}
    \label{fig:7}
\end{figure*}
Untwisting this configuration results in an intermediate triconcave shape characterized by three large outward buckling noses along the prolate. Each of those, as well the three areas in between, can host a vortex. Together with the two initial vortices this gives the maximum of vortices observed in the simulated regime, namely eight. The spurious peaks to nine vortices in Figure \ref{fig:6} are due to a defect merging process, where our algorithm mistakes a saddle point, which merges with two vortices, for a third vortex. Interestingly, those vortices between the noses, where the surface buckles inwards, seem to be stronger than the ones on the outward buckling parts. This is evident by the observation that the latter type of vortex does not always occur and if it does, it is not stable as the evolution proceeds, typically reducing both outward and inward buckling. This reduction of curvature gradients annihilates these vortices by a nearby pair of saddle points, one of which takes the place of the vortex. In particular for smaller $V_r$, the finally obtained shape is stomatochyte-like, often with one bulge bending strongly inwards while the other two bent less strong or flatten out. Much to our surprise, the vortices can even survive this flattening and seem to stay stable on areas without curvature gradients and thus without the odd forces maintaining them. An example showing all of these dynamics is displayed in Figure \ref{fig:7}.

\section{Discussion}
\label{sec:5}

Our numerical study confirms the huge potential of models for fluid deformable surfaces with active geometric forces as an effective approach towards modeling morphogenesis and development. As outlined in \cite{alIzzi} these forces lead to spontaneous shear flows in the presence of mean curvature gradients, which due to the tight coupling of tangential flows and shape deformations can induce shape instabilities. Even though numerical analysis results do not yet exist for this type of equations, the considered realization using surface finite elements \cite{dziuk2013finite,nestler2019finite} provides a reliable framework also for these geometric active systems, leading to presumably optimal convergence rates, which have been established for simplified subproblems.

While our focus here is on the numerical demonstration of these shape instabilities, we would like to make same connection to the targeted application. Addressing the relevant parameters in eqs. \eqref{eq:fds1}-\eqref{eq:fds5} we have $\Reyn = LU\rho/\eta$ and $\Be = L^2U^2\rho/\kappa$, with characteristic length $L$, characteristic velocity $U$, density $\rho$, surface viscosity $\eta$ and bending rigidity $\kappa$. Considering the Scriven-Love number $SL = \Reyn / \Be$ \cite{PhysRevE.101.052401}  allows to eliminate $\rho$. For the other parameters measurements have been done but obviously the values are expected the vary strongly throughout the realm of tissues. As a first assurance we consider $\kappa \sim 1.9 \times 10^{-13}Nm$, measured in \cite{Fouchard_Wyatt_CurlingOfEpithelialMonolayers} and $\eta \sim 10^{-2}\frac{Ns}{m}$, measured via the shear viscosity of the actomyosin cortex \cite{FISCHERFRIEDRICH2016589, Hosseini2023}. Considering typical system sizes and velocities in these experiments lead to $L \sim 2.5 \times 10^{-5}m$ and $U \sim 2.5 \times 10^{-5}\frac{m}{s}$. Putting this together leads to $SL \sim 1.6$, which is within the considered parameter range, for which a shape instability was observed. This further underlines the potential of the approach.

But even if the obtained surface flow fields and shapes are reminiscent to the once seen in C. elegan zygots \cite{10.7554/eLife.04165,bhatnagar2023axis} a quantitative comparison will require additional effects. In our setting the surface area is fixed, which certainly is not the case during development. Changes in surface area in models for fluid deformable surface have already been introduced by adding a source term to eq. \eqref{eq:fds2} \cite{toshniwal2021isogeometric,krause2024wrinkling}. We consider a constant activity $\Ac = 1$, which is unrealistic as well.  Coupling the strength of activity to protein concentrations following from pattern forming systems would therefore be another step towards more quantitative modeling. These are directions for future work. However, already the provided results contain new information on the dynamic coupling of shape changes and surface flow, which ask for experimental validation.\\

{\bf Acknowledgments} \\
M.P. and A.V. were supported by the German Research Foundation (DFG) within grant FOR3013. We further acknowledge computing resources provided at FZ Jülich within grant MORPH and at ZIH/TU Dresden within grant WIR. \\

{\bf Author declaration} \\
Conflict of interest: The authors have no conflicts to disclose. \\
Author contributions: M.P.: formal analysis; software development; simulations; visualization; writing – original draft.
A.V.: conceptualization; methodology; writing – review and editing. \\

{\bf Data availability} \\
Data are available from Zenodo at \href{https://10.5281/zenodo.13333551}{https://10.5281/zenodo.13333551}.

\bibliography{literature.bib}{}

\end{document}